%% file: main_arxiv1.tex
\documentclass[
 reprint,   
 amsmath,amssymb,
 aps,
]{revtex4-2}

\usepackage{comment}
\usepackage{graphicx}
\usepackage{dcolumn}
\usepackage{bm}
\usepackage{comment}

\usepackage{commath}
\usepackage[mathscr]{euscript}
\usepackage{color}
\usepackage{physics}
\usepackage{tensor}
\usepackage{hyperref}
\usepackage{multirow}
\usepackage[dvipsnames]{xcolor}
\usepackage{bclogo}
\usepackage{comment}
\usepackage{dsfont}
\usepackage{cancel}
\usepackage[normalem]{ulem} 
\usepackage{nicefrac}
\usepackage{microtype}
\usepackage{etoolbox}
\usepackage{dsfont}
\usepackage{braket}
\usepackage{mathtools}

\usepackage{tikz}
    \usetikzlibrary{decorations.pathreplacing}
    \usetikzlibrary{decorations.pathmorphing} 
    \usetikzlibrary{decorations.markings} 
    \usetikzlibrary{arrows.meta,bending}
    \usetikzlibrary{decorations}
    \usetikzlibrary{shapes.geometric}
    \usetikzlibrary{shapes}
    \usetikzlibrary{positioning}
    \usetikzlibrary{intersections}
    \usetikzlibrary{external}
    \usetikzlibrary{arrows,patterns, calc,through,backgrounds}
    \usetikzlibrary{shapes.symbols}
\usepackage{tikz-feynman}

\numberwithin{equation}{section}

\newcommand{\FK}{\text{FK}}

\newcommand{\e}[1]{\operatorname{e}^{#1}}

\renewcommand{\d}{\operatorname{d}\!}

\newcommand{\tf}{t_\text{f}}
\newcommand{\ti}{t_\text{i}}
\newcommand{\bdry}{\text{bdry}}
\newcommand{\tmes}[1]{\frac{\d^3 #1}{(2\pi)^3 2 E_{#1}}}
\newcommand{\kmes}[1]{\frac{\d^3 #1}{(2\pi)^3 2 \omega_{#1}}}

\newcommand{\pf}{\ensuremath >}          
\newcommand{\nf}{\ensuremath <}         
\usepackage{xcolor}

\definecolor{RED}{rgb}{1,0,0}

\begin{document}

\preprint{APS/123-QED}

\title{A generating functional for infrared-safe QED amplitudes}

\author{Martin Ammon}
\email{martin.ammon@uni-jena.de}
\author{Konrad Brandts}%
 \email{konrad.brandts@uni-jena.de}
\author{Federico Capone}
\email{federico.capone@uni-jena.de}
 \author{Jakob Hollweck}
\email{jakob.hollweck@uni-jena.de}
\affiliation{%
Theoretisch-Physikalisches Institut \\
Friedrich-Schiller-Universität \\
Fröbelstieg 1, 07743 Jena, Germany
}%

\date{\today}

\begin{abstract}
We construct a generating functional for infrared-finite scattering amplitudes in massive quantum electrodynamics (QED) by including Faddeev-Kulish dressings in a holomorphic coherent-state representation, in the spirit of the construction of Aref'eva-Faddeev-Slavnov (AFS) for the undressed Dyson S-matrix. At tree level, the dressed generating functional reduces to an AFS path integral with dressed boundary conditions. 
\end{abstract}

\maketitle

\section{Introduction}

The S-matrix is central throughout quantum physics, from particle physics to quantum gravity, including the long-standing pursuit of flat-space holography \cite{Polchinski:1999ry}, now developed notably through the \emph{Celestial} and \emph{Carrollian} programmes, e.g. \cite{Zhu:2026ofh,Ruzziconi:2026bix}.

However, the Dyson S-matrix of theories with long-range interactions in four-dimensional Minkowski spacetime is plagued by infrared (IR) singularities. Such singularities are divergences at each order in perturbation theory and, upon resummation, drive all S-matrix elements to zero.

A practical way to circumvent this breakdown of the Haag--Ruelle scattering theory  and to extract observationally meaningful cross sections is to consider only inclusive rates ~\cite{Bloch:1937pw,Kinoshita:1962ur,Lee:1964is,Frye:2018xjj}. However, this does not by itself provide an amplitude-level description
of the long-range interactions responsible for the divergences and obscures the rich infrared structure of gauge theories. Thus, a definition of the S-matrix that is intrinsically free of IR singularities would instead provide both conceptual clarity and technical control.

The prototypical example of an IR-finite S-matrix is the Faddeev--Kulish (FK) construction~\cite{Kulish:1970ut}. In this approach, scattering states are dressed by a soft cloud through a \emph{dressing operator} $\hat{R}_f$ and a \emph{Coulomb phase operator} $\hat{\Phi}$, in such a way that the resulting amplitudes are infrared finite~\footnote{A typical criticism to the FK construction is that it moves the singularity from the amplitude to the asymptotic states. However, as pointed out in \cite{Feal:2022iyn,Feal:2022ufw}, similar criticisms can be moved to any IR-finite S-matrix construction.}.  Despite subtleties in its original derivation within QED \cite{Contopanagos:1991yb}, the construction can be placed on firmer grounds (at least up to certain orders in perturbation theory)  by suitable modifications~\cite{Duch:2021} and it can be shown that the same dressing procedures can be extended to other theories, including linearised gravity, e.g.  \cite{Ware:2013zja}.

In recent years, the FK-type constructions have illuminated the rich IR dynamics of gauge theories mentioned above, which can be characterised in terms of the so-called \emph{ infrared triangles} relating soft factorization theorems, large-gauge symmetries, and memory effects (see~\cite{He:2014cra,Campiglia:2015qka,Gabai:2016kuf,Kapec:2017tkm,Choi:2024mac,Donnay:2026urd} for a partial list of original references). From this viewpoint, the vanishing of the Dyson S-matrix signals its failure to capture transitions between degenerate vacua. In contrast, FK dressed states organise the Hilbert space in superselection sectors labelled by the asymptotic charges, consistently accounting for such vacuum transitions and yielding finite amplitudes. For example, in (linearised) gravity, the soft factorization of amplitudes encodes the asymptotic symmetry structure of spacetime  and is realized in infrared-safe amplitudes of FK type or its generalizations (see e.g. \cite{Choi:2017ylo} and \cite{Prabhu:2022zcr,Prabhu:2024zwl,Prabhu:2024lmg} for related discussions). 

Despite this activity, explicit computations of infrared-dressed amplitudes remain scarce -- a point emphasized in \cite{Hannesdottir:2019opa,Hannesdottir:2019umk} and \cite{Feal:2022iyn,Feal:2022ufw}, which themselves provide some of the few procedures and examples available. 

A second contribution of Faddeev, together with Aref'eva and Slavnov, that is central to this paper is the S-matrix generating functional that bears their names (AFS) \cite{Arefeva:1974jv,Faddeev:1980be}. This  holomorphic coherent-state functional bypasses the Lehmann–Symanzik–Zimmerman (LSZ) reduction formula and is defined as a path integral with non-trivial boundary conditions. This construction (known as \emph{perturbiner formalism} in other contexts, e.g. \cite{Rosly:1996vr,Adamo:2021rfq,LipinskiJusinskas:2026ctz}) has recently been brought back to prominence by a series of works~\cite{Jain:2023fxc,Kim:2023qbl,Kraus:2024gso,Kraus:2025wgi,Isen:2026xoc}, which identify it as a natural object for unifying the study of amplitudes and asymptotic symmetries, as well as a tool to build the holographic dictionary of Carrollian holography \cite{Ammon:2025jmy}.

However, the AFS generating functional generates the Dyson S-matrix  and is thus, by definition, affected by IR-divergences.  This motivates the search for an AFS-like generating functional associated instead with an intrinsically IR-finite S-matrix.

This Letter contributes to this landscape by constructing  an infrared-finite S-matrix generating functional for QED with massive fermions. We call the result, equations \eqref{eq:S_FK_def} and \eqref{eq:pathint}, the \emph{dressed AFS generating functional}, as it extends the original  Aref'eva–Faddeev–Slavnov (AFS) construction applicable solely to the Dyson amplitude. The construction we provide incorporates FK dressings formally at all loops, including the Coulomb phase. Since non-trivial QED scattering changes the asymptotic soft-sector, the fully dressed generating functional naturally encodes this transition and only at tree level reduces to an ordinary AFS path integral with dressed boundary conditions. The relevant terminology is explained in due course. 

The letter is organised as follows. Essential background about coherent states, and notation, are given in section \ref{sec:coherentstates}. The generating functional is introduced and derived in section \ref{sec:genfunc} and its path integral representation in the special case of tree diagrams in section \ref{sec:pathint}. Section \ref{sec:example} discusses Bremsstrahlung as an explicit example. We conclude in section \ref{sec:conclusion}.

\section{QED Coherent states}\label{sec:coherentstates}
We deal with coherent states for Dirac fermions and the gauge field. We define coherent states abstractly in terms of the displacement operator $\hat D$ acting on the Fock space vacuum 
\begin{align}\label{eq:defcoherent}
\begin{split}
\ket{z} = \hat D(z)\ket{0},\qquad \ket{z}_N=\mathrm{e}^{-|z|^2/2}\ket{z}
\end{split}
\end{align}
where only the state $\ket{z}_N$ is normalised and $z \in \mathbb{C}$. In the case of fermions created by $\hat{b}^\dagger_s$ and $\hat{d}^\dagger_s$ ($s=1,2$) satisfying, $\{\hat{b}_r(\mathbf{p}), \hat{b}_s^\dagger(\mathbf{p}')\}=\{\hat{d}_r(\mathbf{p}), \hat{d}_s^\dagger(\mathbf{p}')\}= (2\pi)^3 2 E_{\mathbf{p}}\, \delta_{rs} \delta^{(3)}(\mathbf{p}-\mathbf{p}')$,
the displacement operator reads 
\begin{align} \label{eq:fermion_displacement_operator}
&\hat D_b(z_1)\coloneqq \exp\left(\int\widetilde{\mathrm{d}^3 \mathbf{p}}\, \hat b^\dagger_s(\mathbf{p}) z_{1}^s(\mathbf{p}) \right),
\end{align}
with a similar expression for $D_d(z_2)$ associated to $d_s^\dagger$ with eigenvalue $z_2^s$. Here, $z_{1,2}^s$ and their barred counterparts are independent Grassmann variables. We denote here $\widetilde{\mathrm{d}^3 \mathbf{p}}\coloneqq \tmes{\mathbf p}$ and we sum over spin-indices, $s$.

For the gauge field $A_\mu$,  we work in a covariant gauge and introduce the polarisation vectors $\epsilon_\mu^{(\lambda)}$, satisfying $\epsilon^{(\lambda)}\cdot\epsilon^{(\lambda')}=\eta^{\lambda\lambda'}$ and $\sum_{\lambda,\lambda'=0}^3\eta_{\lambda\lambda'}\epsilon^{(\lambda)}_\mu \epsilon^{(\lambda')}_\nu =\eta_{\mu\nu}$, with $\eta$ in mostly plus signature.  The annihilation operators decompose as $\hat a_\mu= \epsilon_\mu^{(\lambda)} \hat a_\lambda$ and $\hat a_\lambda=\eta_{\lambda\lambda'}\epsilon^{(\lambda')\mu}\hat a_\mu$, using the Einstein summation convention. The commutation relations are  $[\hat a_\mu(\mathbf{k}),\hat a_\nu^\dagger(\mathbf{k}')] = (2\pi)^3 2 \omega_{\mathbf{k}} \eta_{\mu \nu} \delta^{(3)}(\mathbf{k}-\mathbf{k'})$, or equivalently in polarisation basis $[\hat a_\lambda(\mathbf{k}),\hat a_{\lambda'}^\dagger(\mathbf{k}')]
=(2\pi)^3 2\omega_{\mathbf{k}}\eta_{\lambda\lambda'}
\delta^{(3)}(\mathbf{k}-\mathbf{k}')$.  We define the displacement operator  as
\begin{equation} \label{eq:operator_photon_displacement}
\hat D_a(z_\gamma)\coloneqq\exp\left(\int\widetilde{\mathrm{d}^3 \mathbf{k}}\,\hat a_\lambda^\dagger(\mathbf{k}) z^\lambda(\mathbf{k})\right), 
\end{equation}
where $\widetilde{\mathrm{d}^3 \mathbf{k}}\coloneqq \kmes{\mathbf k}$. Here and in the following, we denote dependence on photon coherent states by the $\gamma$-subscript (and drop it where it is superfluous because of indices).

Let us list a few properties of the coherent states defined in equations \eqref{eq:fermion_displacement_operator} and \eqref{eq:operator_photon_displacement}. For this, we define functional derivatives in fermionic and bosonic coherent states by, for example 
\begin{equation}
\overset{\rightarrow}{\delta}_{\bar z_1^s(\mathbf{p})}:= (2\pi)^3 2 E_\mathbf{p} \frac{\overset{\rightarrow}{\delta}}{\delta \bar z_1^s(\mathbf{p})}\,,\quad \delta_{z_\mu(\mathbf{k})}:= (2\pi)^3 2 \omega_\mathbf{k} \dfrac{\delta}{\delta z^\mu(\mathbf{k})}\,,
\end{equation}
where we defined $\dfrac{\delta}{\delta z^\mu(\mathbf{k})}=\epsilon^{(\lambda)}_\mu(\mathbf{k)}\dfrac{\delta}{\delta z^\lambda(\mathbf{k})}$. These derivative operators encode the action of creation operators on the functional space as 
\begin{align}
 \ket{z_1} \overset{\leftarrow}{\delta}_{z_1^s(\mathbf{p})} = \hat{b}_{s} ^\dagger (\mathbf{p}) \ket{z_1}\,,&\quad  \overset{\rightarrow}{\delta}_{\bar{z}_1^s(\mathbf{p})} \bra{\bar{z}_1} = \bra{\bar{z}_1}\hat{b}_{s} (\mathbf{p}) \,, \notag \\  \delta_{z^\mu(\mathbf{k})} \ket{z_\gamma} &= \hat{a}_{\mu} ^\dagger (\mathbf{k}) \ket{z_\gamma}\,. \label{eq:coherent_state_derivative_prop}
\end{align}
The coherent states define an overcomplete, non-orthogonal basis. We write for photons: 
\begin{subequations} \label{eq:notation_CS}
\begin{align}
    \langle \bar{z}_\gamma | z'_\gamma \rangle &= \exp\left(\int \widetilde{d^3\mathbf{k}}\,  \bar{z}^\lambda (\mathbf{k}) z'_{\lambda}(\mathbf{k})\right) =: \mathrm{e}^{(\bar z_\gamma,z'_\gamma)} \label{eq:notation_inner_prod} \\
    \mathds{1} &= \int \mathcal{D}\bar{z}_\gamma\mathcal{D}z_\gamma \mathrm{e}^{- \int \widetilde{\mathrm{d}^3\mathbf{k}} \bar{z}^\lambda(\mathbf{k}) z_\lambda(\mathbf{k}) } \ket{z_\gamma}\!\!\!{}\bra{\bar{z}_\gamma}\,, \\
   \mathcal{D}\bar{z}_\gamma\mathcal{D}z_\gamma &= \left(\prod_{\mathbf{k},\lambda} \frac{\mathrm{d}\bar{z}
   _\lambda(\mathbf{k}) \mathrm{d}z_\lambda(\mathbf{k})}{2\pi i}\right)\,.
\end{align}
\end{subequations}
Note that we introduced an alternative notation for the overlap $\langle \bar{z}_\gamma | z'_\gamma \rangle$ in equation \eqref{eq:notation_inner_prod} which is at some points of the main text more convenient. The equations \eqref{eq:notation_CS} hold analogously for the fermionic coherent states.

Finally, a general QED coherent state is then generated by displacement operators of electrons, positrons and photons. For example, the displacement operator associated to a general in-going QED coherent state $\ket{\beta}$ is given by
\begin{equation}
\hat D(\beta) = \hat D_{b}(\beta_1)  \hat D_d(\beta_2) \hat D_a(\beta_\gamma)\,,
\end{equation}
where here and in the following we denote all outgoing coherent states by the label $\bar \alpha$ and all in-going ones by $\beta$.

\section{The generating functional}\label{sec:genfunc}
The Faddeev-Kulish amplitude is defined by \cite{Kulish:1970ut} 
\begin{align}\label{eq:FKamplitude}
&\mathcal{M}_{m,n}^{\text{FK}}\\ &=  \lim_{\substack{|t|  \to \infty \\ \lambda_s \to 0}} \braket{p_1,\dots ,p_m|e^{\hat{R}_f} e^{-i \hat \Phi(t_f)} \hat S(t) e^{i \hat \Phi(t_i)}e^{-\hat{R}_f}  |p_1',\dots,p_n'}\,.\notag
\end{align}
The labels  $m$ and $n$ denote the number of outgoing and ingoing particles respectively (we suppress dependence on spin and polarization labels).  $\hat S$ is the Dyson S-operator expressed  via the evolution operator $\hat U$ and the free evolution operator $\hat{U}_0$, $ \hat S(t) \coloneqq 
    \hat U_0^\dagger(\tf, t_0)\hat U(\tf, \ti)\hat U_0(\ti, t_0)$,
and we condense the dependence on $(t_f,t_i)$ in the letter $t$. $\hat R_f$ is the dressing operator
\begin{align} \label{eq:def_R_operator}
    \hat R_f = e \int \widetilde{\mathrm{d}^3 \mathbf{k}} \int \widetilde{\mathrm{d}^3 \mathbf{p}} \left(f^\mu(p, k) \hat a^\dagger_\mu(\mathbf k) - h.c.\right) \hat \rho(\mathbf p)
\end{align}
where $h.c.$ denotes the Hermitian conjugate ($\bar f^\mu$ for $f^\mu$) and 
\begin{align}
    &\hat \rho(\mathbf p) = \hat d^\dagger_s(\mathbf p) \hat d^s(\mathbf p) - \hat b^\dagger_s(\mathbf p)\hat b^s(\mathbf p)\,, \\
    &f^\mu(p, k) = \left(\frac{p^\mu}{p\cdot k} - c^\mu(k)\right)\phi(p, k)\,, \label{eq:def_f}
\end{align}
with $c^\mu(k)$ being a null vector such that $c\cdot k =1$ and $\lim_{\omega_{\mathbf k}\to 0} \phi(p, k) = 1$. Note that the choice of $\phi$ is not fixed by the requirement of IR cancellations. The operator $\hat R_f$ can be defined similarly to \cite{Hirai:2022yqw} in a window $\lambda_s\leq \omega_{\mathbf k}\leq\Lambda_s$ (which we do not write explicitly) where $\lambda_s$ is the infrared cutoff and $\Lambda_s$ is a fixed scale delimiting the soft region. To keep the notation concise, we generally suppress the explicit dependence of the objects below on these regulators, as it will be clear from context which objects depend on them. Whenever we discuss the limit $\lambda_s\to 0$ we intend to take it after all computations are performed at finite $\Lambda_s$ and after the large time limit, so that the soft pole developed in the Dyson amplitude is appropriately cancelled.   
$\hat\Phi(t)$ is the Coulomb phase operator~\footnote{Note that the literature, including the original computation in \cite{Kulish:1970ut}, uses mostly $:\hat{\rho}(\mathbf p) \hat{\rho}(\mathbf q):$ instead, the difference being self-energy terms. This means we assume here instead that UV divergences in the theory have been renormalized, but without getting rid of the IR divergences of the self-energy diagrams. Note, however, that the formalism works either way.} 
\begin{align} \label{eq:def_Phi_operator}
    \hat\Phi(t) = \frac{e^2}{8\pi} \int \widetilde{\mathrm{d}^3\mathbf{p}} \int \widetilde{\mathrm{d}^3\mathbf{q}}\frac{p \cdot q}{  \sqrt{(p \cdot q)^2 - m^4}} \hat{\rho}(\mathbf p) \hat{\rho}(\mathbf q) \int^t \frac{\mathrm{d}\tau}{|\tau|}\,. 
\end{align}
Our first main result is a generating functional for these amplitudes. Before presenting it, some remarks are in order. The amplitude is IR-finite because the dressing operator corresponds to that of Chung \cite{Chung:1965zza}. However, the FK derivation from the asymptotic dynamics is rather cavalier. See \cite{Contopanagos:1991yb,Duch:2021} and \cite{Hirai:2022yqw} for a derivation of \eqref{eq:def_R_operator} from appropriate limiting procedures of time-dependent operators $\hat{R}(t)$. Motivated by the evidence that the FK prescription yields the correct infrared-finite dressed amplitude (see, e.g., \cite{Hannesdottir:2019opa}), we bypass this step and directly construct the coherent-state generating functional corresponding to the amplitude above.

Given coherent states $\ket{\alpha}$ and $\ket{\beta}$, and the AFS generating functional $\mathcal{S}[\bar\alpha, \beta;t]$ of the undressed Dyson amplitude, we define the \emph{dressed AFS generating functional} via the finite action of differential operators on $\mathcal{S}$,
\begin{align} \label{eq:S_FK_def}
    &\mathcal{S}^\text{FK}[\bar \alpha, \beta;t]\\ &\coloneqq e^{-i \overset{\rightarrow}{\Phi}_{\text{out}}(\bar\alpha,t_f)} e^{ \overset{\rightarrow}{R}_{\text{out}}(\bar\alpha)}  \mathcal{S}[\bar\alpha, \beta;t] e^{- \overset{\leftarrow}{R}_{\text{in}}(\beta)} e^{i \overset{\leftarrow}{\Phi}_{\text{in}}(\beta,t_i)} \,.\notag\
\end{align}
Derivatives of the undressed AFS functional with respect to the coherent states data $\bar\alpha$ and $\beta$ generates the Fock-space amplitudes. Similarly, the dressed momentum-space amplitudes are obtained by differentiation of \eqref{eq:S_FK_def} with respect to  the coherent state data, as \footnote{When polarization or spin indices are omitted, the derivatives or generators of the Heisenberg algebra correspond to either the photon or the fermion algebra.}
\begin{align}\label{eq:result}
\mathcal{M}_{m,n}^{\text{FK}}= \lim_{\substack{|t|  \to \infty \\ \lambda_s \to 0 }} \bigg[\prod_{j=1}^{m}\overset{\rightarrow}{\delta}_{\bar\alpha(\mathbf q_j)} \mathcal{S}^\text{FK}[\bar \alpha, \beta;t] \prod_{i=1}^{n} \overset{\leftarrow}{\delta}_{\beta(\mathbf p_i)} \bigg]_{\bar\alpha, \beta = 0}\,.
\end{align}
As for equation \eqref{eq:FKamplitude}, we stress that both the large time limit, and the selection of the Fock sector through functional derivatives have to be taken before removing the soft regulator $\lambda_s$. Under the assumption that at finite $\lambda_s$ the finite time amplitudes converge to the corresponding FK-amplitudes, one may take the large time limit before selecting a sector. Notice that, while the FK amplitude can be evaluated by construction directly between fixed Fock states, the generating functional is a natural object in coherent state basis.

Proving \eqref{eq:result} is immediate. Let $\mathcal{O}$ be any operator and define its coherent-state kernel by \begin{align}
    \mathcal{O}[\bar{\alpha},\beta] \coloneqq \bra{\bar\alpha}\mathcal O\ket{\beta}\,.
\end{align}
The derivative operator acts on the coherent states as in \eqref{eq:coherent_state_derivative_prop} and returns scattering states in momentum basis upon setting the coherent state data $\bar\alpha$ and $\beta$ to zero, namely thanks to the defining property \eqref{eq:defcoherent}. For example, 
\begin{align}
    \prod_{i=1}^n  \ket{\beta} \overset{\leftarrow}{\delta}_{\beta(\mathbf{p}_i)}\bigg|_{\beta = 0} = \prod_{i=1}^n \hat b^\dagger(\mathbf{p}_i) \ket{0} = \ket{p_1,\cdots,p_n}\,,
\end{align}
and similarly for outgoing states. Consequently, for the FK-dressed operator 
\begin{align}\label{eq:s_fk_def}
    \hat{S}^{\text{FK}}(t) \coloneqq  \mathrm{e}^{\hat R_f} \mathrm{e}^{-i\hat\Phi(t_f)} 
    \hat S(t)
      \mathrm{e}^{i\hat\Phi(t_i)} \mathrm{e}^{-\hat R_f}
\end{align}
one obtains equation \eqref{eq:result} up to the identification \eqref{eq:S_FK_def}, which is obtained in the following way. With standard manipulations using equation \eqref{eq:coherent_state_derivative_prop}, we can write
\begin{align}
  \e{i\hat \Phi}  \e{-\hat R_f}  \ket{\beta} =  \ket{\beta}   \e{i\overset{\leftarrow}{\Phi}(\beta)} \e{-\overset{\leftarrow}{R}_f(\beta)}\,,
\end{align}
with a similar expression for the outgoing states. This effectively turns the dressing operators into derivative operators that can be moved outside of the inner product
\begin{align} \label{eq:def_densities}
\begin{split}
    \overset{\rightarrow}{\rho}_{\mathrm{out}} &= \bar{\alpha}^s_2(\mathbf{p}) \overset{\rightarrow}{\delta}_{\bar{\alpha}^s_2(\mathbf{p})} - \bar{\alpha}^s_1(\mathbf{p}) \overset{\rightarrow}{\delta}_{\bar{\alpha}^s_1(\mathbf{p})} \\
    \overset{\leftarrow}{\rho}_{\mathrm{in}} &=  \overset{\leftarrow}{\delta}_{\beta^s_2(\mathbf{p})} \beta^s_2(\mathbf{p}) -  \overset{\leftarrow}{\delta}_{\beta^s_1(\mathbf{p})} \beta^s_1(\mathbf{p})\,,
    \end{split}
\end{align}
directly yielding the result \eqref{eq:S_FK_def}.

Let us now evaluate the operators in equation \eqref{eq:S_FK_def} to find a more explicit form of $\mathcal{S}^\text{FK}[\bar \alpha, \beta;t]$ and discuss its constituents. To this end, let us take for the moment only $\mathcal{S}[\bar{\alpha},\beta;t]$. The operators $\overset{\rightarrow}{\rho}_{\mathrm{out}}$ and $\overset{\leftarrow}{\rho}_{\mathrm{in}}$ in the Coulomb phase operator and dressing operator probe for the fermion coherent states to which an individual cloud of photons is then added. Therefore, it is natural to resolve the generating functional $\mathcal{S}^{\text{FK}}[\bar{\alpha},\beta;t]$ into fermionic sectors by expanding in fermionic sources and evaluating the fermionic differential operators 
    \begin{widetext}
    \begin{align} \label{eq:dressing_expansion}
    &\mathcal{S}^{\text{FK}}[\bar\alpha;\beta;t]
    \\ &=
    \sum_{m_1,m_2,n_1,n_2}
    \frac{\int [\mathrm{d} \mathbf q]^{m_1}[\mathrm{d}\mathbf{q}]^{m_2}[\mathrm{d}\mathbf p]^{n_1}[\mathrm{d}\mathbf{p}]^{n_2}}{m_1!m_2!n_1!n_2!}
    \,
    \bar\alpha_1^{m_1}\bar\alpha_2^{m_2} \mathrm{e}^{-i\Phi_{\mathrm{out}}(t_f)}\ e^{ \overset{\rightarrow}{R}_{\text{out,h}}(\bar\alpha_\gamma)} \mathcal M_{m_1,m_2;n_1,n_2}[\bar \alpha_\gamma,\beta_\gamma;t] e^{- \overset{\leftarrow}{R}_{\text{in},h}(\beta_\gamma)} \mathrm{e}^{i\Phi_{\mathrm{in}}(t_i)}\
    \beta_1^{n_1}\beta_2^{n_2}\,, \notag
\end{align}
\end{widetext}
where $\bar{\alpha}_1^{m_1} = \prod_{i=1}^{m_1} \bar{\alpha}_{1,s_i}(\mathbf q_i)$, $[\mathrm d \mathbf q]^{m_1} := \prod_{i=1}^{m_1}\widetilde{\mathrm{d}^3 \mathbf q_i}$ and analogously for the other parameters. This is convenient because the densities \eqref{eq:def_densities} act diagonally on a fixed fermionic sector in the expansion. The Coulomb phase operator is completely evaluated this way
\begin{gather}  \label{eq:coulomb_and_eval_densities}
    \mathrm{e}^{-i\overset{\rightarrow}{\Phi}_{\mathrm{out}}}\,\bar{\alpha}_1^{m_1}\bar{\alpha}_2^{m_2}
    =:\mathrm{e}^{-i\Phi_{\mathrm{out}}}\,\bar{\alpha}_1^{m_1}\bar{\alpha}_2^{m_2} \\
    \rho_{\mathrm{out}}(\mathbf q)
=
    \sum_{a=1}^{m_2}\tilde \delta^{(3)}(\mathbf q-\mathbf{q}_a)
    -
    \sum_{b=1}^{m_1}\tilde \delta^{(3)}(\mathbf{q}-\mathbf{q}_b)
\end{gather}
where we have defined the evaluated phase operator $\Phi_{\text{out}}$ through the evaluated density operator $\rho_{\mathrm{out}}$ and \eqref{eq:def_Phi_operator}; this is again done in the same way for the in-variables. We furthermore have defined $\tilde \delta^{(3)}(\mathbf p) = (2\pi)^3 2 E_{\mathbf p} \delta( \mathbf p)$. The operators $ \overset{\rightarrow}{R}_{\text{out,h}}$ and $\overset{\leftarrow}{R}_{\text{in,h}}$ are the dressing operators after evaluation of the densities in equation \eqref{eq:def_densities}. Note that they depend explicitly on the charged sector, while at the same time being operators in the photon sector. The outgoing dressing operator is for example given by 
\begin{align} \label{eq:def_diff_R_and_F}
     \overset{\rightarrow}{R}_{\text{out,h}} &= e \int \widetilde{\mathrm{d}^3 \mathbf k} \left(\bar \alpha_\mu(\mathbf k)F_{\text{out}}^\mu(\mathbf k)  - \bar{F}_{\text{out},\mu}(\mathbf k) \overset{\rightarrow}{\delta}_{\bar \alpha_\mu(\mathbf k)}\right) \notag\\
    F_{\text{out}}^\mu(\mathbf k) &= \int \widetilde{\mathrm{d}^3 \mathbf p} \,\rho_{\text{out}}(\mathbf p) f^\mu(p,k)\,.
\end{align}
Both dressing operators are still understood as being regulated by $\lambda_s$.

The conjugation in photon variables in equation \eqref{eq:dressing_expansion} can then further be evaluated by recognizing that $\mathrm{e}^{\overset{\rightarrow}{R}_{\text{out,h}}}$ and $\mathrm{e}^{\overset{\leftarrow}{R}_{\text{in},h}}$ are differential coherent state displacement operators similar~\footnote{Note that these differential displacement operators depend on $\bar \alpha_\mu$; calling this also a displacement operator is subject to notation.} to equation \eqref{eq:operator_photon_displacement} such that (we drop the indices of $\mathcal{M}$ in the following)
\begin{align} \label{eq:conj_photon_kernel}
   & e^{ \overset{\rightarrow}{R}_{\text{out,h}}(\bar\alpha_\gamma)} \mathcal M[\bar \alpha_\gamma,\beta_\gamma] e^{- \overset{\leftarrow}{R}_{\text{in},h}(\beta_\gamma)}\\ & = \mathcal{N}_{\text{oo}} \mathrm{e}^{e(\bar\alpha_\gamma,F_{\text{out}})} \mathcal{M}[\bar{\alpha}'_{\gamma}, \beta'_\gamma] \mathrm{e}^{e(\bar{F}_{\text{in}},\beta_\gamma)}\mathcal{N}_{\text{ii}} \notag 
\end{align}
for the shifted coherent state labels $\bar{\alpha}'_{\mu} = \bar \alpha_\mu-e\bar{F}_{\text{out},\mu}$, $\beta'_\mu =\beta_\mu - eF_{\text{in},\mu}$ and where we used the notation introduced in \eqref{eq:notation_inner_prod}.  $\mathcal{N}_{\text{oo}}$ and $\mathcal{N}_{\text{ii}}$ are, respectively, the normalization factors coming from cloud overlaps between out-out and in-in states
\begin{align}
    \mathcal{N}_{\text{oo}} = \mathrm{e}^{-\frac{e^2}{2} (\bar F_{\text{out}},F_{\text{out}})}\,, \quad \mathcal{N}_{\text{ii}} = \mathrm{e}^{-\frac{e^2}{2} (\bar F_{\text{in}},F_{\text{in}})}\,.
\end{align}

It is now convenient to separate the normal symbol $\mathcal{T}[\bar{\alpha}'_{\gamma},\beta'_\gamma]$ of $\mathcal{M}[\bar{\alpha}'_{\gamma}, \beta'_\gamma]$ from the kernel of free photon propagation (i.e. for $\hat S = \mathds{1}$), 
$\langle \bar \alpha_\gamma | \beta_\gamma \rangle = \mathrm{e}^{(\bar{\alpha}_\gamma,\beta_\gamma)}$, as
\begin{align} \label{eq:T_kernel}
    \mathcal{M}[\bar{\alpha}'_{\gamma}, \beta'_\gamma] = \mathrm{e}^{(\bar{\alpha}'_\gamma,\beta'_\gamma)} \mathcal{T}[\bar{\alpha}'_{\gamma}, \beta'_\gamma]\,,
\end{align}
The normal symbol $\mathcal{T}$ is the part of $\mathcal{M}$ generated by interactions \cite{Kraus:2025wgi}. The free-photon kernel is completely factored out from the non-trivial part $\mathcal{T}$ and only contributes to the coherent state boundary factors. However, it contributes through the shift as
\begin{align}
    \mathrm{e}^{(\bar{\alpha}'_\gamma,\beta'_\gamma)} =  \mathrm{e}^{(\bar \alpha_\gamma, \beta_\gamma)}  \mathrm{e}^{-e(\bar \alpha_\gamma, F_{\text{in}})} \mathrm{e}^{-e(\bar F_{\text{out}}, \beta_\gamma)} \mathrm{e}^{e^2(\bar F_{\text{out}}, F_{\text{in}})}\,.
\end{align}
Defining the one-particle cloud factors for each external fermion
\begin{align}
\begin{split}
    W_{\text{in}}(f) &:= \mathrm{e}^{-e(\bar \alpha_\gamma, F_{\text{in}})} \mathrm{e}^{e(\bar F_{\text{in}},\beta_\gamma)}\,, \\ W_{\text{out}}(f) &:=  \mathrm{e}^{e(\bar \alpha_\gamma,F_{\text{out}})} \mathrm{e}^{-e(\bar F_{\text{out}}, \beta_\gamma)} 
\end{split}
\end{align}
and 
\begin{equation}
    \mathcal{N}_{\text{oi}} = \mathrm{e}^{e^2(\bar F_{\text{out}}, F_{\text{in}})}\,,
\end{equation}
accounting for the overlap between incoming and outgoing clouds (discussed in more detail in section \ref{sec:pathint}), the dressed photon kernel \eqref{eq:conj_photon_kernel} can be decomposed as
\begin{align} \label{eq:eval_conj_kernel}
\begin{split}
    & e^{ \overset{\rightarrow}{R}_{\text{out,h}}(\bar\alpha_\gamma)} \mathcal M[\bar \alpha_\gamma,\beta_\gamma] e^{- \overset{\leftarrow}{R}_{\text{in},h}(\beta_\gamma)}\\
    &= \mathrm{e}^{(\bar \alpha_\gamma, \beta_\gamma)} \mathcal{N}\, W_{\text{in}}(f) W_{\text{out}}(f) \mathcal{T}[\bar{\alpha}'_{\gamma},\beta'_\gamma]\,,
\end{split}
\end{align}
where $\mathcal{N}=\mathcal{N}_{\text{oo}}  \mathcal{N}_{\text{oi}} \mathcal{N}_{\text{ii}}={}_N \langle F_{\text{out}}| F_{\text{in}}\rangle_N$ is the normalized coherent state overlap of $F_{\text{in}}$ and $F_{\text{out}}$, 
This overlap is independent of the photon coherent-state variables $\bar\alpha$ and $\beta$ and is fixed by the hard-sector data entering the corresponding dressings. When the infrared regulator is removed, $\lambda_s \to 0$ at fixed $\Lambda_s$, $\mathcal{N}$ vanishes if the in- and out-clouds fall in different superselection sectors (their total leading soft profiles differ).

Equations \eqref{eq:dressing_expansion} and \eqref{eq:eval_conj_kernel} are therefore the explicit form of the full dressed AFS generating functional $\mathcal{S}^{\text{FK}}$ with all operators evaluated and deserve some interpretative words.

First, equations \eqref{eq:eval_conj_kernel} and \eqref{eq:dressing_expansion} reflect directly that the imaginary divergences of the Dyson $S$-matrix (according to the classification in \cite{Weinberg:1965nx}) only appear for loops involving either in-in our out-out legs. Moreover, the real divergences appear for all combinations of in- and out-legs. Most importantly, the divergences stemming from combinations of in- and out-going particles are cancelled by the cloud-overlap factors.

Second, the appearance of the Coulomb phase operator $\Phi$ is very natural from the perspective of the photon-cloud sector. To see this, consider the time dependent dressing operator $\hat R(t)$ in \cite{Kulish:1970ut}. After evaluating the charge densities, it defines a charge-dependent displacement $\hat R_h(t)$ in the photon labels. Coherent states, however, carry also phase information. Thus, the lift of this displacement to the actual coherent state space contains an additional scalar phase, which is precisely the Coulomb phase.\footnote{In the language of non-cyclic geometric phases, the Coulomb phase is therefore a dynamical phase \cite{PhysRevA.52.2576}. For the normalized photon coherent states one may write it in terms of the Berry connection one-form associated with the cloud photon coherent state space. For example in the quantum-mechanical case, $A = i{}_N \langle z|d|z \rangle_N$, where $\ket{z}_N$ is the normalised coherent state in \eqref{eq:defcoherent} and $d$ the exterior derivative. With our sign convention, $\Phi = -\int A$.} Equivalently, since the commutator  $[\hat R_h(t_2),\hat R_h(t_1)] =: -i\sigma(t_2,t_1)\hat {\mathds 1}$ is central, the composition of $\mathrm{e}^{\hat R_h(t_2)}$ and $\mathrm{e}^{\hat R_h(t_1)}$ closes only up to $\mathrm{e}^{-\frac{i}{2}\sigma(t_2,t_1)}$. The Coulomb phase is therefore the accumulated version of this central phase $\sigma$ along the time-dependent FK path; infinitesimally $\dot \Phi(t) = \frac{i}{2}[R_h(t),\dot R_h(t)]$. Hence, at finite time the Coulomb phase is intrinsic to the photon displacement that defines the FK dressing.\footnote{Note that this differs from the non-relativistic case, where only the Coulomb phase contributes.} When using the time-independent dressing $R_f$, $\Phi$ should still be understood as the accumulated central phase of the underlying time-dependent displacement. This discussion motivates why the Coulomb phase should not be formally neglected, although it does not contribute to cross sections. Some other crucial reasons can be found for example in \cite{Lippstreu:2025jit}.\footnote{Furthermore, in gravity, the Coulomb phase is related to classical observables such as the Shapiro time delay.}

\section{Path integral  description at tree level}\label{sec:pathint}

The undressed AFS generating functional, $\mathcal{S}[\bar{\alpha},\beta;t]$ can be expressed as a path integral with non-trivial boundary conditions, satisfying a variational principle where the positive frequency modes of a free field are fixed in the far past and the negative frequency modes of the field are fixed in the far future \cite{Faddeev:1980be,Kim:2023qbl}. The dressed AFS generating functional cannot be represented by the same simple, local, sector-independent AFS boundary description because the boundary data would explicitly depend on the charged sector of the chosen scattering process via the dependence of the evaluated dressing operator and the Coulomb phase on $F_{\text{in}}$ and $F_{\text{out}}$. However, we show that for tree level computations things simplify and the dressed AFS generating functional \eqref{eq:dressing_expansion} reduces to an AFS path integral with dressed boundary conditions. 

Below, we organize the perturbative coefficients of \eqref{eq:dressing_expansion} in terms of connected QED Feynman diagrams and strip off the trivial free overlap. 

Consider an amplitude with $E_f$ external fermions, $E_\gamma$ external photons and $V$ vertices. The number of loops for a connected graph is given by $L = \frac12(V - E_f - E_\gamma) + 1$. The different contributions to \eqref{eq:dressing_expansion} contribute the following orders of the coupling constant, $e$. Both the Coulomb phases $\mathrm{e}^{\pm i \Phi_{\text{in,out}}}$ and the normalization factors $\mathcal{N}_{\text{oo}}  \mathcal{N}_{\text{oi}}  \mathcal{N}_{\text{ii}}$ are exponentials with arguments independent of coherent states and of order $\mathcal{O}(e^2)$. Hence, they do not contribute photon or fermion legs to order $L$, but only starting from order $L' = L+1$.  Similarly, expanding $\mathcal{T}[\bar{\alpha}'_{\gamma},\beta'_\gamma]$ in \eqref{eq:T_kernel} in photon coherent states, one checks that the shift only contributes from order $L+1$. Thus, these terms do not contribute at tree level. Maybe more intuitively, as we have discussed in the last section, this discards (at finite $\lambda_s$) the geometric overlap and phase information relating different FK clouds and leaves only the cloud contributions corresponding to individual particles.

Hence, the dressing contributes to tree level diagrams only through $W_{\text{in}}(f)$ and $ W_{\text{out}}(f)$. We can redistribute them among the fermions $\bar\alpha_1^{m_1}\bar\alpha_2^{m_2}$ and $ \beta_1^{n_1}\beta_2^{n_2}$, using equations \eqref{eq:coulomb_and_eval_densities} and \eqref{eq:def_diff_R_and_F}, in the expansion \eqref{eq:dressing_expansion} 
    \begin{widetext}
    \begin{align} \label{eq:dressing_expansion_tree}
    \mathcal{S}^{\text{FK}}_{\text{tree}}[\bar\alpha;\beta;t]
    =
    \sum_{m_1,m_2,n_1,n_2}
    \frac{\int [\mathrm{d} \mathbf q]^{m_1}[\mathrm{d} \mathbf q]^{m_2}[\mathrm{d}\mathbf p]^{n_1}[\mathrm{d}\mathbf p]^{n_2}}{m_1!m_2!n_1!n_2!}
    \,
   \bar \alpha_{1,W}^{m_1}\bar\alpha_{2,W}^{m_2}   \mathcal M_{m_1,m_2;n_1,n_2}[\bar \alpha_\gamma,\beta_\gamma;t]
    \beta_{1,W}^{n_1}\beta_{2,W}^{n_2} = \mathcal S[\bar \alpha_W,\beta_W;t]\,, 
\end{align}
\end{widetext}
where for example
\begin{align}
    \beta_{1,W}^{n_1} &:= \prod_{i=1}^{n_1} \beta_{1,s_i}(\mathbf q_i)W_{\text{in},1}(f, \mathbf q_i)\,,\\
    W_{\text{in},1}(f, \mathbf q_i) &:= \mathrm{e}^{e (\bar \alpha_\gamma,f(\mathbf q_i)) }\mathrm{e}^{- e(\bar f( \mathbf q_i),\beta_\gamma)}\,. \label{eq:1p_dressing}
\end{align}
The chain of equalities in  \eqref{eq:dressing_expansion_tree} states that at tree level the dressed AFS generating functional, $\mathcal{S}^{\text{FK}}_{\text{tree}}[\bar\alpha;\beta;t]$ is an AFS path integral with dressed boundary conditions, which we denote by $\mathcal{S}[\bar \alpha_W,\beta_W,t]$ (at tree level).

We can express the latter as 
\begin{equation}\label{eq:pathint}
 \mathcal{S}^{\text{FK}}_{\text{tree}}[\varphi_i,\varphi_f]
    =  \smashoperator{\int_{\varphi_i,\varphi_f}} \mathcal D \bar\psi\mathcal D\psi \mathcal D A \e{i\left(I[\bar \psi, \psi, A] + I_\bdry[\bar \psi, \psi, A; \varphi_i,\varphi_f]\right)}\,,
\end{equation}
where the right hand side is evaluated in the semi-classical limit. The integration in equation \eqref{eq:pathint} is over spinor and gauge fields $\psi,\,\bar \psi,\,A_\mu$; $I$ is the QED action in Feynman gauge and $I_\bdry$ are boundary terms required to make the action well defined with the chosen boundary conditions ($\varphi_i,\varphi_f$) at early and late times ($t_i, t_f$). Labelling with $>$ the positive frequency part of a field and with $<$ its negative frequency part, the appropriate boundary conditions consist of dressing and keeping fixed these parts of the Dirac field as
\begin{align}\label{eq:dressed_afs_pi_boundary_conditions}
        \psi_>(\ti, \mathbf x)     & = \sum_s \int \widetilde{\dd^3 \mathbf p}\, u_s(p) W_{\text{in},1}(f,\mathbf p) \beta_1^s(\mathbf p) \mathrm{e}^{ip \cdot x}\\
         \psi_<(\tf, \mathbf x)     & = \sum_s \int \widetilde{\dd^3 \mathbf p}\, v_s(p) W_{\text{out},2}(f,\mathbf p) \bar \alpha_2^s(\mathbf p) \mathrm{e}^{-ip \cdot x} \nonumber
 \end{align}
 with similar formulae for $\bar \psi$. At the same time,  $A^\mu_<(\tf, \mathbf x) $, $A^\mu_>(\ti, \mathbf x) $ are fixed but remain undressed.  In the path integral we denote the set of boundary fields at $t_i$ collectively with $\varphi_i=\{\psi_>(\ti, \mathbf x),\bar \psi_>(\ti, \mathbf x),A^\mu_>(\ti, \mathbf x)\}$ and similarly those at $t_f$ with $\varphi_f$.  Note that $W_{\text{in},1}(f,\mathbf p) = W_{\text{out},2}(f,\mathbf p)$ for the dressings defined in \eqref{eq:1p_dressing}.
 
 The corresponding boundary terms are given by 
 \begin{align} \label{eq:I_bdry}
\begin{split}
 I_\bdry = &-(\bar \psi_\nf, \psi )_{\Sigma_f} + (\bar\psi_\pf, \psi )_{\Sigma_i}\\
 &+ (A^\mu_\nf, A_\mu )_{\Sigma_f} - (A^\mu_\pf, A_\mu )_{\Sigma_i}\,, 
\end{split}
\end{align}
where $(\cdot,\cdot)$ denotes the inner products
 \begin{gather} \label{eq:inner_products}
    (\bar\psi, \psi )_\Sigma = \int_\Sigma \dd^3 x\; \bar \psi \gamma^0 \psi \\ 
   (A^\mu, B_\mu )_\Sigma = \frac{1}{2}\int_\Sigma \dd^3 x\;  (A^\mu \dot B_\mu - \dot A^\mu B_\mu ) \notag
 \end{gather}
defined on a Cauchy slice $\Sigma$. Notice that in a generic gauge, additional boundary terms would appear.

This concludes our proofs and shows that at tree level, the dressing is part of the prescribed boundary data which is held fixed under variation of the bulk fields. The variational principle is the same as in the undressed AFS construction \cite{Kim:2023qbl,  Kraus:2024gso, Kraus:2025wgi,Isen:2026xoc}, but with boundary fields replaced by dressed boundary fields. 
In the large-time limit this expression, which depends on the soft regulator,  generates the IR-finite result as we show in the next section.

\section{An example}\label{sec:example}
\begin{figure}[t]
\centering
\input{figure1_new}
\caption{ Electron-line contributions to the radiative current. The electron legs and the photon of momentum $q$ are on shell. The unrestricted momentum $k=p_1-p_2-q$ enters through a current insertion. The two panels denote the two orderings of soft-photon emission from the electron line.}
\label{fig:bremsstrahlung_clouds}
\end{figure}
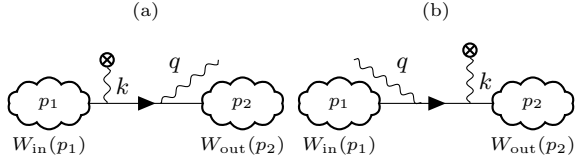

As a proof of concept, we compute the tree-level radiative electron current with
on-shell electrons, $p_1^2=p_2^2=-m^2$, one on-shell photon, $q^2=0$ and
$q^0=\omega_{\mathbf{q}}>0$, and an unrestricted current momentum
$k=p_1-p_2-q$. In particular no mass-shell condition is imposed on $k$. This object is not a stand-alone S-matrix process. It is a radiative current
that can be embedded, for example, in electron scattering from a heavy charged
target, which is also known to be Bremsstrahlung. In a complete target amplitude its open current index is contracted
with a conserved target current, and for target emission and target absorption dressing must
also be included.

\begin{figure}[t]
\centering
\input{figure2_new}
\caption{Cloud contributions to the dressed radiative current. The on-shell soft photon of momentum $q$ is emitted from the incoming cloud in $(a)$ and from the outgoing cloud in $(b)$. In both panels, the unrestricted momentum $k$ enters at the current insertion. The full cloud-emission contribution is the sum of both diagrams.
}
\label{fig:cloud_soft_emission}
\end{figure}
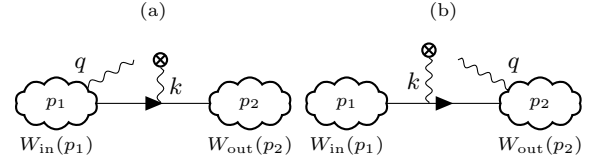

It is customary in this context to use Feynman diagrams modified by clouds that attach to the in- and out-going fermions (see e.g. \cite{Kapec:2017tkm}). The two electron-line orderings are depicted in Fig.~\ref{fig:bremsstrahlung_clouds}. We show that the dressed generating functional $\mathcal{S}^{\text{FK}}_{\text{tree}}$ in Eq.~\eqref{eq:pathint} generates their leading soft pole together with the cloud-emission terms in Fig.~\ref{fig:cloud_soft_emission}.\footnote{The factors we discarded in section \eqref{sec:pathint} to stay at tree level are diagrammatically represented by loops with integration over soft momenta. The normalizations $\mathcal{N}_{\text{ii}}$, $\mathcal{N}_{\text{oo}}$ and $\mathcal{N}_{\text{oi}}$ contribute photon lines that connect different in- and out-going clouds. The Coulomb-phase is usually denoted as a line between two Fermion legs and the shift corresponds to the remaining possible processes, a soft photon line connecting a cloud with the bulk QED graph. }

To generate the perturbative expansion, we need the free coherent-state kernel on which the interaction operator acts. Let $\mathcal S_0[\bar\alpha_W,\beta_W;\bar\eta,\eta,J]$ denote the free AFS kernel for QED with dressed boundary data. Its normal symbol is then given by $\mathcal{S}_0^N$ 
\begin{equation}
    \mathcal{S}_0^N = \mathrm{e}^{-(\bar \alpha_W,\beta_W)} \mathcal{S}_0[\bar \alpha_W,\beta_W;\bar \eta,\eta,J]\,.
\end{equation}
Accounting only for the connected non-trivial part of \eqref{eq:pathint}, we construct the interactions through the action of $\mathrm{e}^{iV}$ as 
\begin{align}
    \mathcal{T} := \mathcal{T}^{\text{FK}}_{\text{tree}} = (\mathrm{e}^{iV} \mathcal{S}_0^N[\bar \alpha_W,\beta_W;\bar \eta,\eta,J])\Big|_{\eta = \bar \eta = J = 0}\,,
\end{align}
with
\begin{align}
    V[\delta J, \delta \eta, \delta \bar \eta] = -ie \int \mathrm{d}^4x \overset{\rightarrow}{\delta}_{\bar{\eta}^\alpha(x)} (\gamma^\mu)^\alpha{}_{\beta} \overset{\leftarrow}{\delta}_{\eta_\beta(x)} \delta_{J^\mu(x)}\,.
\end{align}
To access an unrestricted current momentum, introduce a classical bulk probe
$B_\mu$ directly in the interaction,
\begin{equation}
 I_{\rm int}[A,B]
 =-e\int d^4x\,\bar\psi\gamma^\mu\psi\,(A_\mu+B_\mu).
 \label{eq:interaction_external_current}
\end{equation}We write $\mathcal T[B]$ for the normal symbol obtained by making this
replacement in the same perturbative expansion.

Taking functional derivatives w.r.t. $B_\mu(x)$ inserts the current 
\begin{equation}
 \left.\frac{\delta}{\delta B_\mu(x)}e^{iI[A,B]}\right|_{B=0}
 =-ie\,\bar\psi(x)\gamma^\mu\psi(x)e^{iI[A,0]}.
 \label{eq:B_identity_external_current}
\end{equation}

Expanding up to the orders relevant for the radiative current coefficient, we define
\begin{align}
\mathcal{T}^{(1)}
&=
-ie \int d^4x\,
\mathcal{A}_\mu(x)\,
\bar{\psi}_\alpha(x)\,
(\gamma^\mu)^\alpha{}_\beta\,
\psi^\beta(x),
\\[0.5em]
\mathcal{T}^{(2)}
&=
\frac{(-ie)^2}{2}\int \mathrm{d}^4x \mathrm{d}^4y \notag\\
&\begin{aligned}[t]
\Big[
&\bar{\psi}_\beta(y)\,
K^\beta{}_\alpha(y,x;\nu,\mu)\,
\psi^\alpha(x)\,
\mathcal{A}_\nu(y) \mathcal{A}_\mu(x)
\\
&+
\bar{\psi}_\beta(x)\,
K^\beta{}_\alpha(x,y;\mu,\nu)\,
\psi^\alpha(y)\,
\mathcal{A}_\mu(x) \mathcal{A}_\nu(y)
\Big]
\end{aligned}
\end{align}
for $K^\beta{}_\alpha(y,x;\nu,\mu)
:=
(\gamma^\nu)^\beta{}_\gamma\,
S_F(y-x)^\gamma{}_\delta\,
(\gamma^\mu)^\delta{}_\alpha$ and $S_F$ the Feynman propagator. Here, $\mathcal{A}_\mu := A_\mu + B_\mu.$ The radiative current with momentum $k$ is inserted by acting with the operator
\begin{equation}
 \mathfrak D_B^\mu(k):=
 \int \mathrm{d}^4x\,\mathrm{e}^{-ik\cdot x}\frac{\delta}{\delta B_\mu(x)}
 \label{eq:DB_external_current}
\end{equation}

In analogy to \eqref{eq:result}, we define then the dressed radiative current at finite $q$ as 
\begin{align}
 \mathcal H_{\FK}^{\mu\rho}(k,q)
 :=
 \mathfrak D_B^\mu(k)
 \delta_{\bar\alpha_\rho(\mathbf q)}
 \overset{\rightarrow}{\delta}_{\bar\alpha_{s_2}(\mathbf p_2)}\notag\\
 \qquad\times\mathcal{T}[B]
 \overset{\leftarrow}{\delta}_{\beta_{s_1}(\mathbf p_1)}
 \Big|_{\bar\alpha=\beta=B=0}.
 \label{eq:radiative_current_definition}
\end{align}
The index $\mu$ belongs to the off-shell current insertion. The index $\rho$
belongs to the on-shell photon and is contracted with
$\epsilon_\rho^{(\sigma)}(q)$. The relevant terms of $\mathcal{T}[B]$ are given by $\mathcal{T}[B] = \mathcal{T}^{(1)} + \mathcal{T}^{(2)}$. 
Accordingly, we write $\mathcal H_{\FK}^{\mu\rho}(k,q)= \mathcal H^{\mu\rho \, (1)}(k,q) + \mathcal H^{\mu\rho \, (2)}(k,q)$ 
in the following. The hard amplitude is given by
\begin{align}
    \mathcal H^\mu(k_0) =-ie\,\bar u^{s_2}(p_2)\gamma^\mu u^{s_1}(p_1)
\end{align}
where $k_0 = p_1-p_2$ and $k_{0\mu}\mathcal H^\mu(k_0)=0.$ Moreover, we  compute
\begin{align}
   \mathcal H^{\mu\rho \, (2)}(k,q)
&=P^\rho(p_1,p_2,q)\mathcal H^\mu(k_0)
 +O(\omega_{\textbf{q}}^{0}),
\label{eq:bulk_soft_external_current}\\
P^\rho(p_1,p_2,q)
 &=e\left(\frac{p_2^\rho}{p_2\cdot q}
 -\frac{p_1^\rho}{p_1\cdot q}\right).
 \label{eq:P_external_current}
\end{align}
This contribution is represented by the two diagrams in figure \ref{fig:bremsstrahlung_clouds}. The shift $k-k_0=-q$ affects the finite term but not the leading pole.
 
In the dressed case, one derivative may act on $\mathcal T^{(1)}$, producing the hard photon, while the other acts on the FK dressing, supplying the additional soft photon from either the incoming or outgoing cloud, as in Fig. \ref{fig:cloud_soft_emission}:
\begin{align} \label{eq:one_vertex}
- S^\rho(p_1,p_2, q)W_{\text{in}}(f,\mathbf{p}_1)W_{\text{out}}(f,\mathbf{p}_2) \mathcal{H}^\mu
\end{align}
for 
\begin{align}
    S^\rho(p_1,p_2,q) &= e(f^\rho(p_2,q) - f^\rho(p_1,q))
\end{align}
with $f$ defined in \eqref{eq:def_f}. In the example of the radiative current, $W_{\text{in}}(f,\mathbf{p}_1)W_{\text{out}}(f,\mathbf{p}_2)$ contribute at order $e^2$ only with their constant term, such that 
\begin{align} 
\mathcal H^{\mu\rho \, (1)}(k,q) = -S^\rho (p_1,p_2,q)  \mathcal{H}^\mu(k_0)\,.
\end{align}
This contribution is represented by the two diagrams in figure \ref{fig:cloud_soft_emission}, where we want to stress that the clouds depict the asymptotic states as source for soft photons, additionally to the bulk vertex. Using the definition \eqref{eq:def_f} and taking $\phi(p,q) = 1$, one finds that $S^\mu(q) = P^\mu(q)$, where the $c_\mu$ have been cancelled using charge conservation. Therefore, the leading soft coefficient vanishes at tree level
\begin{align} 
   \lim_{\omega_{\mathbf{q}} \to 0} \omega_{\mathbf{q}} \mathcal H_{\FK}^{\mu\rho}(k,q)= 0\,.
\end{align}
This completes the tree-level example, the coherent state functional correctly generated the amplitude without the $\omega^{-1}_{\mathbf{q}}$ pole. Equation
\eqref{eq:radiative_current_definition} applies the general extraction
prescription \eqref{eq:result}, with the additional derivative
$\mathfrak D_B^\mu(k)$ inserting the external current. The coherent-state
derivatives continue to select only on-shell asymptotic particles.

The example shows that operator insertions can be incorporated into the dressed AFS formalism by supplementing the coherent-state derivatives with appropriate source derivatives, without modifying the FK dressing  prescription.

We omit a higher-order check because the cancellation of the virtual IR divergences at higher orders is well-established in the operator formalism \cite{Chung:1965zza} and we have demonstrated via \eqref{eq:dressing_expansion} how to reproduce the corresponding terms in the context of the dressed AFS generating functional. Such a computation would naturally use the full dressed AFS generating functional in \eqref{eq:dressing_expansion} at finite time and soft regulator (removed after taking the defining derivatives of the corresponding amplitude), and an expansion of $\mathcal{T}[\bar{\alpha}'_{\gamma}, \beta'_\gamma]$ in photon coherent states.

\section{Conclusions}\label{sec:conclusion}

In this Letter we construct a generating functional for FK-dressed QED amplitudes, motivated by the AFS formalism, which we consider to be particularly natural for the purpose because it bypassess the LSZ reduction procedure. We  derive a holomorphic generating functional valid at all loop orders and evaluate it explicitly (section \ref{sec:genfunc}) and we  show  that at tree level it admits an AFS path-integral representation with FK-dressed boundary data (section \ref{sec:pathint}).

Pragmatically, we use the FK amplitude with  asymptotic dynamics partially resolved in terms of the operator $\hat R_f$, thus avoiding the time-dependent dressing $\hat R(t)$. While this is common, it would be interesting to extend the holomorphic formalism to adiabatic constructions of the modified S-matrix \cite{Duch:2021,Hirai:2022yqw} (established up to a relatively low perturbative order), the algebraic approach of \cite{Prabhu:2024lmg}, or alternative dressings such as \cite{Choi:2026swo}.

We expect our construction to connect naturally with the study of asymptotic symmetries and flat-space holography. Extensions to linearised gravity \cite{Isen:2026xoc} and subleading soft dressings \cite{Choi:2019rlz,Christodoulou:2026gvt}, as well as links to soft effective actions \cite{Choi:2026dhz} and the space of soft vacua \cite{Kapec:2022hih}, would be particularly interesting. It would also be useful to clarify the asymptotic behaviour of the fields corresponding to the dressed boundary conditions and understand their representation-theoretical content at null infinity \cite{Donnay:2026urd}.

The relation between the AFS path integral and the quantum effective action also merits further study. The undressed AFS path integral can be interpreted as the quantum effective action evaluated on solutions of the effective equations of motion with prescribed asymptotic data \cite{Jevicki:1987ax}. We expect this to extend to IR-dressed states, reproducing our result at tree level, with Wilson-line representations of the dressings likely playing a key role.

In conclusion, we view our contribution as a timely complement to existing approaches to constructing explicit IR-safe observables. While we do not claim that our approach supersedes \cite{Hannesdottir:2019opa,Hannesdottir:2019umk} or \cite{Feal:2022iyn,Feal:2022ufw}, we believe this perspective is worth developing further.

\section*{Acknowledgements}

We thank P. Kraus for insightful discussions, M. Campiglia and R. Myers for comments on a draft and related discussions, and S. Agrawal, G. Barnich, S. Choi, S. Christodoulou, L. Donnay, V. Nenmeli, and C. Sieling for useful exchanges.  FC is supported by the Deutsche Forschungsgemeinschaft (DFG) under Grant No 406116891 within the Research Training Group RTG 2522/1. JH is funded by a \emph{Landesgraduiertenstipendium} of the federal state of Thuringia.

\bibliography{bib}

\end{document}

%% file: figure1_new.tex

\tikzset{
    dressedcloud/.style={
        cloud,
        cloud puffs=10,
        cloud puff arc=120,
        aspect=2,
        draw,
        thick,
        fill=white,
        minimum width=0.75cm,
        minimum height=0.10cm
    },
    crossedcircle/.style={
        circle,
        draw,
        thick,
        inner sep=0pt,
        minimum size=5pt,
        path picture={
            \draw[thick]
                (path picture bounding box.south west)
                -- (path picture bounding box.north east);
            \draw[thick]
                (path picture bounding box.north west)
                -- (path picture bounding box.south east);
        }
    }
}

\begin{tikzpicture}[scale=0.64, baseline=(current bounding box.center)]

\node at (-3.0,1.9) {\scriptsize (a)};

\begin{feynman}
\vertex (Lpi)   at (-5.00, 0.00);
\vertex (LVh)   at (-3.8, 0);
\vertex (LVs)   at (-2.7, 0);
\vertex (Lpf)   at (-1.00, 0);
\vertex[crossedcircle] (Lk) at (-3.8, 0.90) {};
\vertex (Lq)    at (-1.5, 0.90);

\diagram*{
  (Lpi) -- [fermion] (Lpf),
  (LVh) -- [photon, edge label'=\(k\)] (Lk),
  (LVs) -- [photon, edge label=\(q\)] (Lq),
};
\end{feynman}

\node[dressedcloud] at (Lpi) {\scriptsize $p_1$};
\node[dressedcloud] at (Lpf) {\scriptsize $p_2$};
\node[below] at (-5.0,-0.50) {\scriptsize $W_{\mathrm{in}}(p_1)$};
\node[below] at (-1.0,-0.50) {\scriptsize $W_{\mathrm{out}}(p_2)$};

\node at (3.0,1.9) {\scriptsize (b)};

\begin{feynman}
\vertex (Rpi)   at (1, 0.0);
\vertex (RVs)   at (2.7, 0.0);
\vertex (RVh)   at (3.7, 0.0);
\vertex (Rpf)   at (5, 0.0);
\vertex (Rq)    at (1.30, 0.9);
\vertex[crossedcircle] (Rk) at (3.7, 1.1) {};

\diagram*{
  (Rpi) -- [fermion] (Rpf),
  (RVs) -- [photon, edge label'=$q$] (Rq),
  (RVh) -- [photon, edge label'=$k$] (Rk),
};
\end{feynman}

\node[dressedcloud] at (Rpi) {\scriptsize $p_1$};
\node[dressedcloud] at (Rpf) {\scriptsize $p_2$};
\node[below] at (1, -0.5) {\scriptsize $W_{\mathrm{in}}(p_1)$};
\node[below] at (5, -0.5) {\scriptsize $W_{\mathrm{out}}(p_2)$};

\end{tikzpicture}

%% file: figure2_new.tex

\tikzset{
    dressedcloud/.style={
        cloud,
        cloud puffs=10,
        cloud puff arc=120,
        aspect=2,
        draw,
        thick,
        fill=white,
        minimum width=0.75cm,
        minimum height=0.10cm
    },
    crossedcircle/.style={
        circle,
        draw,
        thick,
        inner sep=0pt,
        minimum size=5pt,
        path picture={
            \draw[thick]
                (path picture bounding box.south west)
                -- (path picture bounding box.north east);
            \draw[thick]
                (path picture bounding box.north west)
                -- (path picture bounding box.south east);
        }
    }
}

\begin{tikzpicture}[scale=0.64, baseline=(current bounding box.center)]

\node at (-3.0,1.9) {\scriptsize (a)};

\begin{feynman}
\vertex (Lpi)   at (-5.00, 0.00);
\vertex (LVh)   at (-2.85, 0); 
\vertex (LVs)   at (-5, 0);    
\vertex (Lpf)   at (-1.00, 0);
\vertex[crossedcircle] (Lk) at (-2.85, 0.90) {};
\vertex (Lq)    at (-3.4, 0.90);

\diagram*{
  (Lpi) -- [fermion] (Lpf),
  (LVh) -- [photon, edge label'=$k$] (Lk),
  (LVs) -- [photon, edge label=$q$] (Lq),
};
\end{feynman}

\node[dressedcloud] at (Lpi) {\scriptsize $p_1$};
\node[dressedcloud] at (Lpf) {\scriptsize $p_2$};
\node[below] at (-5.0,-0.50) {\scriptsize $W_{\mathrm{in}}(p_1)$};
\node[below] at (-1.0,-0.50) {\scriptsize $W_{\mathrm{out}}(p_2)$};

\node at (3.0,1.9) {\scriptsize (b)};

\begin{feynman}
\vertex (Rpi)   at (1, 0.0);
\vertex (RVs)   at (5, 0.0);   
\vertex (RVh)   at (2.7, 0.0); 
\vertex (Rpf)   at (5, 0.0);
\vertex (Rq)    at (3.30, 0.9);
\vertex[crossedcircle] (Rk) at (2.7, 1.1) {};

\diagram*{
  (Rpi) -- [fermion] (Rpf),
  (RVs) -- [photon, edge label'=\(q\)] (Rq),
  (RVh) -- [photon, edge label=\(k\)] (Rk),
};
\end{feynman}

\node[dressedcloud] at (Rpi) {\scriptsize $p_1$};
\node[dressedcloud] at (Rpf) {\scriptsize $p_2$};
\node[below] at (1, -0.5) {\scriptsize $W_{\mathrm{in}}(p_1)$};
\node[below] at (5, -0.5) {\scriptsize $W_{\mathrm{out}}(p_2)$};

\end{tikzpicture}

%% file: main_arxiv1.bbl
\begin{thebibliography}{55}%
\makeatletter
\providecommand \@ifxundefined [1]{%
 \@ifx{#1\undefined}
}%
\providecommand \@ifnum [1]{%
 \ifnum #1\expandafter \@firstoftwo
 \else \expandafter \@secondoftwo
 \fi
}%
\providecommand \@ifx [1]{%
 \ifx #1\expandafter \@firstoftwo
 \else \expandafter \@secondoftwo
 \fi
}%
\providecommand \natexlab [1]{#1}%
\providecommand \enquote  [1]{``#1''}%
\providecommand \bibnamefont  [1]{#1}%
\providecommand \bibfnamefont [1]{#1}%
\providecommand \citenamefont [1]{#1}%
\providecommand \href@noop [0]{\@secondoftwo}%
\providecommand \href [0]{\begingroup \@sanitize@url \@href}%
\providecommand \@href[1]{\@@startlink{#1}\@@href}%
\providecommand \@@href[1]{\endgroup#1\@@endlink}%
\providecommand \@sanitize@url [0]{\catcode `\\12\catcode `\$12\catcode `\&12\catcode `\#12\catcode `\^12\catcode `\_12\catcode `\%12\relax}%
\providecommand \@@startlink[1]{}%
\providecommand \@@endlink[0]{}%
\providecommand \url  [0]{\begingroup\@sanitize@url \@url }%
\providecommand \@url [1]{\endgroup\@href {#1}{\urlprefix }}%
\providecommand \urlprefix  [0]{URL }%
\providecommand \Eprint [0]{\href }%
\providecommand \doibase [0]{https://doi.org/}%
\providecommand \selectlanguage [0]{\@gobble}%
\providecommand \bibinfo  [0]{\@secondoftwo}%
\providecommand \bibfield  [0]{\@secondoftwo}%
\providecommand \translation [1]{[#1]}%
\providecommand \BibitemOpen [0]{}%
\providecommand \bibitemStop [0]{}%
\providecommand \bibitemNoStop [0]{.\EOS\space}%
\providecommand \EOS [0]{\spacefactor3000\relax}%
\providecommand \BibitemShut  [1]{\csname bibitem#1\endcsname}%
\let\auto@bib@innerbib\@empty
\bibitem [{\citenamefont {Polchinski}(1999)}]{Polchinski:1999ry}%
  \BibitemOpen
  \bibfield  {author} {\bibinfo {author} {\bibfnamefont {J.}~\bibnamefont {Polchinski}},\ }\bibfield  {title} {\bibinfo {title} {{S matrices from AdS space-time}},\ }\href@noop {} {\  (\bibinfo {year} {1999})},\ \Eprint {https://arxiv.org/abs/hep-th/9901076} {arXiv:hep-th/9901076} \BibitemShut {NoStop}%
\bibitem [{\citenamefont {Zhu}(2026)}]{Zhu:2026ofh}%
  \BibitemOpen
  \bibfield  {author} {\bibinfo {author} {\bibfnamefont {B.}~\bibnamefont {Zhu}},\ }\bibfield  {title} {\bibinfo {title} {{Topics in Celestial holography: A bottom-up perspective}},\ }\href@noop {} {\  (\bibinfo {year} {2026})},\ \Eprint {https://arxiv.org/abs/2606.24285} {arXiv:2606.24285 [hep-th]} \BibitemShut {NoStop}%
\bibitem [{\citenamefont {Ruzziconi}(2026)}]{Ruzziconi:2026bix}%
  \BibitemOpen
  \bibfield  {author} {\bibinfo {author} {\bibfnamefont {R.}~\bibnamefont {Ruzziconi}},\ }\bibfield  {title} {\bibinfo {title} {{Carrollian physics and holography}},\ }\href {https://doi.org/10.1016/j.physrep.2026.03.005} {\bibfield  {journal} {\bibinfo  {journal} {Phys. Rept.}\ }\textbf {\bibinfo {volume} {1182}},\ \bibinfo {pages} {1} (\bibinfo {year} {2026})},\ \Eprint {https://arxiv.org/abs/2602.02644} {arXiv:2602.02644 [hep-th]} \BibitemShut {NoStop}%
\bibitem [{\citenamefont {Bloch}\ and\ \citenamefont {Nordsieck}(1937)}]{Bloch:1937pw}%
  \BibitemOpen
  \bibfield  {author} {\bibinfo {author} {\bibfnamefont {F.}~\bibnamefont {Bloch}}\ and\ \bibinfo {author} {\bibfnamefont {A.}~\bibnamefont {Nordsieck}},\ }\bibfield  {title} {\bibinfo {title} {{Note on the Radiation Field of the electron}},\ }\href {https://doi.org/10.1103/PhysRev.52.54} {\bibfield  {journal} {\bibinfo  {journal} {Phys. Rev.}\ }\textbf {\bibinfo {volume} {52}},\ \bibinfo {pages} {54} (\bibinfo {year} {1937})}\BibitemShut {NoStop}%
\bibitem [{\citenamefont {Kinoshita}(1962)}]{Kinoshita:1962ur}%
  \BibitemOpen
  \bibfield  {author} {\bibinfo {author} {\bibfnamefont {T.}~\bibnamefont {Kinoshita}},\ }\bibfield  {title} {\bibinfo {title} {{Mass singularities of Feynman amplitudes}},\ }\href {https://doi.org/10.1063/1.1724268} {\bibfield  {journal} {\bibinfo  {journal} {J. Math. Phys.}\ }\textbf {\bibinfo {volume} {3}},\ \bibinfo {pages} {650} (\bibinfo {year} {1962})}\BibitemShut {NoStop}%
\bibitem [{\citenamefont {Lee}\ and\ \citenamefont {Nauenberg}(1964)}]{Lee:1964is}%
  \BibitemOpen
  \bibfield  {author} {\bibinfo {author} {\bibfnamefont {T.~D.}\ \bibnamefont {Lee}}\ and\ \bibinfo {author} {\bibfnamefont {M.}~\bibnamefont {Nauenberg}},\ }\bibfield  {title} {\bibinfo {title} {{Degenerate Systems and Mass Singularities}},\ }\href {https://doi.org/10.1103/PhysRev.133.B1549} {\bibfield  {journal} {\bibinfo  {journal} {Phys. Rev.}\ }\textbf {\bibinfo {volume} {133}},\ \bibinfo {pages} {B1549} (\bibinfo {year} {1964})}\BibitemShut {NoStop}%
\bibitem [{\citenamefont {Frye}\ \emph {et~al.}(2019)\citenamefont {Frye}, \citenamefont {Hannesdottir}, \citenamefont {Paul}, \citenamefont {Schwartz},\ and\ \citenamefont {Yan}}]{Frye:2018xjj}%
  \BibitemOpen
  \bibfield  {author} {\bibinfo {author} {\bibfnamefont {C.}~\bibnamefont {Frye}}, \bibinfo {author} {\bibfnamefont {H.}~\bibnamefont {Hannesdottir}}, \bibinfo {author} {\bibfnamefont {N.}~\bibnamefont {Paul}}, \bibinfo {author} {\bibfnamefont {M.~D.}\ \bibnamefont {Schwartz}},\ and\ \bibinfo {author} {\bibfnamefont {K.}~\bibnamefont {Yan}},\ }\bibfield  {title} {\bibinfo {title} {{Infrared Finiteness and Forward Scattering}},\ }\href {https://doi.org/10.1103/PhysRevD.99.056015} {\bibfield  {journal} {\bibinfo  {journal} {Phys. Rev. D}\ }\textbf {\bibinfo {volume} {99}},\ \bibinfo {pages} {056015} (\bibinfo {year} {2019})},\ \Eprint {https://arxiv.org/abs/1810.10022} {arXiv:1810.10022 [hep-ph]} \BibitemShut {NoStop}%
\bibitem [{\citenamefont {Kulish}\ and\ \citenamefont {Faddeev}(1970)}]{Kulish:1970ut}%
  \BibitemOpen
  \bibfield  {author} {\bibinfo {author} {\bibfnamefont {P.~P.}\ \bibnamefont {Kulish}}\ and\ \bibinfo {author} {\bibfnamefont {L.~D.}\ \bibnamefont {Faddeev}},\ }\bibfield  {title} {\bibinfo {title} {{Asymptotic conditions and infrared divergences in quantum electrodynamics}},\ }\href {https://doi.org/10.1007/BF01066485} {\bibfield  {journal} {\bibinfo  {journal} {Theor. Math. Phys.}\ }\textbf {\bibinfo {volume} {4}},\ \bibinfo {pages} {745} (\bibinfo {year} {1970})}\BibitemShut {NoStop}%
\bibitem [{Note1()}]{Note1}%
  \BibitemOpen
  \bibinfo {note} {A typical criticism to the FK construction is that it moves the singularity from the amplitude to the asymptotic states. However, as pointed out in \cite {Feal:2022iyn,Feal:2022ufw}, similar criticisms can be moved to any IR-finite S-matrix construction.}\BibitemShut {Stop}%
\bibitem [{\citenamefont {Contopanagos}\ and\ \citenamefont {Einhorn}(1992)}]{Contopanagos:1991yb}%
  \BibitemOpen
  \bibfield  {author} {\bibinfo {author} {\bibfnamefont {H.~F.}\ \bibnamefont {Contopanagos}}\ and\ \bibinfo {author} {\bibfnamefont {M.~B.}\ \bibnamefont {Einhorn}},\ }\bibfield  {title} {\bibinfo {title} {{Theory of the asymptotic S matrix for massless particles}},\ }\href {https://doi.org/10.1103/PhysRevD.45.1291} {\bibfield  {journal} {\bibinfo  {journal} {Phys. Rev. D}\ }\textbf {\bibinfo {volume} {45}},\ \bibinfo {pages} {1291} (\bibinfo {year} {1992})}\BibitemShut {NoStop}%
\bibitem [{\citenamefont {Duch}(2021)}]{Duch:2021}%
  \BibitemOpen
  \bibfield  {author} {\bibinfo {author} {\bibfnamefont {P.}~\bibnamefont {Duch}},\ }\bibfield  {title} {\bibinfo {title} {Infrared problem in perturbative quantum field theory},\ }\href {https://doi.org/10.1142/S0129055X2150032X} {\bibfield  {journal} {\bibinfo  {journal} {Reviews in Mathematical Physics}\ }\textbf {\bibinfo {volume} {33}},\ \bibinfo {pages} {2150032} (\bibinfo {year} {2021})},\ \Eprint {https://arxiv.org/abs/1906.00940} {arXiv:1906.00940 [math-ph]} \BibitemShut {NoStop}%
\bibitem [{\citenamefont {Ware}\ \emph {et~al.}(2013)\citenamefont {Ware}, \citenamefont {Saotome},\ and\ \citenamefont {Akhoury}}]{Ware:2013zja}%
  \BibitemOpen
  \bibfield  {author} {\bibinfo {author} {\bibfnamefont {J.}~\bibnamefont {Ware}}, \bibinfo {author} {\bibfnamefont {R.}~\bibnamefont {Saotome}},\ and\ \bibinfo {author} {\bibfnamefont {R.}~\bibnamefont {Akhoury}},\ }\bibfield  {title} {\bibinfo {title} {{Construction of an asymptotic S matrix for perturbative quantum gravity}},\ }\href {https://doi.org/10.1007/JHEP10(2013)159} {\bibfield  {journal} {\bibinfo  {journal} {JHEP}\ }\textbf {\bibinfo {volume} {10}},\ \bibinfo {pages} {159}},\ \Eprint {https://arxiv.org/abs/1308.6285} {arXiv:1308.6285 [hep-th]} \BibitemShut {NoStop}%
\bibitem [{\citenamefont {He}\ \emph {et~al.}(2014)\citenamefont {He}, \citenamefont {Mitra}, \citenamefont {Porfyriadis},\ and\ \citenamefont {Strominger}}]{He:2014cra}%
  \BibitemOpen
  \bibfield  {author} {\bibinfo {author} {\bibfnamefont {T.}~\bibnamefont {He}}, \bibinfo {author} {\bibfnamefont {P.}~\bibnamefont {Mitra}}, \bibinfo {author} {\bibfnamefont {A.~P.}\ \bibnamefont {Porfyriadis}},\ and\ \bibinfo {author} {\bibfnamefont {A.}~\bibnamefont {Strominger}},\ }\bibfield  {title} {\bibinfo {title} {{New Symmetries of Massless QED}},\ }\href {https://doi.org/10.1007/JHEP10(2014)112} {\bibfield  {journal} {\bibinfo  {journal} {JHEP}\ }\textbf {\bibinfo {volume} {10}},\ \bibinfo {pages} {112}},\ \Eprint {https://arxiv.org/abs/1407.3789} {arXiv:1407.3789 [hep-th]} \BibitemShut {NoStop}%
\bibitem [{\citenamefont {Campiglia}\ and\ \citenamefont {Laddha}(2015)}]{Campiglia:2015qka}%
  \BibitemOpen
  \bibfield  {author} {\bibinfo {author} {\bibfnamefont {M.}~\bibnamefont {Campiglia}}\ and\ \bibinfo {author} {\bibfnamefont {A.}~\bibnamefont {Laddha}},\ }\bibfield  {title} {\bibinfo {title} {{Asymptotic symmetries of QED and Weinberg{\textquoteright}s soft photon theorem}},\ }\href {https://doi.org/10.1007/JHEP07(2015)115} {\bibfield  {journal} {\bibinfo  {journal} {JHEP}\ }\textbf {\bibinfo {volume} {07}},\ \bibinfo {pages} {115}},\ \Eprint {https://arxiv.org/abs/1505.05346} {arXiv:1505.05346 [hep-th]} \BibitemShut {NoStop}%
\bibitem [{\citenamefont {Gabai}\ and\ \citenamefont {Sever}(2016)}]{Gabai:2016kuf}%
  \BibitemOpen
  \bibfield  {author} {\bibinfo {author} {\bibfnamefont {B.}~\bibnamefont {Gabai}}\ and\ \bibinfo {author} {\bibfnamefont {A.}~\bibnamefont {Sever}},\ }\bibfield  {title} {\bibinfo {title} {{Large gauge symmetries and asymptotic states in QED}},\ }\href {https://doi.org/10.1007/JHEP12(2016)095} {\bibfield  {journal} {\bibinfo  {journal} {JHEP}\ }\textbf {\bibinfo {volume} {12}},\ \bibinfo {pages} {095}},\ \Eprint {https://arxiv.org/abs/1607.08599} {arXiv:1607.08599 [hep-th]} \BibitemShut {NoStop}%
\bibitem [{\citenamefont {Kapec}\ \emph {et~al.}(2017)\citenamefont {Kapec}, \citenamefont {Perry}, \citenamefont {Raclariu},\ and\ \citenamefont {Strominger}}]{Kapec:2017tkm}%
  \BibitemOpen
  \bibfield  {author} {\bibinfo {author} {\bibfnamefont {D.}~\bibnamefont {Kapec}}, \bibinfo {author} {\bibfnamefont {M.}~\bibnamefont {Perry}}, \bibinfo {author} {\bibfnamefont {A.-M.}\ \bibnamefont {Raclariu}},\ and\ \bibinfo {author} {\bibfnamefont {A.}~\bibnamefont {Strominger}},\ }\bibfield  {title} {\bibinfo {title} {{Infrared Divergences in QED, Revisited}},\ }\href {https://doi.org/10.1103/PhysRevD.96.085002} {\bibfield  {journal} {\bibinfo  {journal} {Phys. Rev. D}\ }\textbf {\bibinfo {volume} {96}},\ \bibinfo {pages} {085002} (\bibinfo {year} {2017})},\ \Eprint {https://arxiv.org/abs/1705.04311} {arXiv:1705.04311 [hep-th]} \BibitemShut {NoStop}%
\bibitem [{\citenamefont {Choi}\ \emph {et~al.}(2025)\citenamefont {Choi}, \citenamefont {Laddha},\ and\ \citenamefont {Puhm}}]{Choi:2024mac}%
  \BibitemOpen
  \bibfield  {author} {\bibinfo {author} {\bibfnamefont {S.}~\bibnamefont {Choi}}, \bibinfo {author} {\bibfnamefont {A.}~\bibnamefont {Laddha}},\ and\ \bibinfo {author} {\bibfnamefont {A.}~\bibnamefont {Puhm}},\ }\bibfield  {title} {\bibinfo {title} {{The classical super-phaserotation infrared triangle. Classical logarithmic soft theorem as conservation law in (scalar) QED}},\ }\href {https://doi.org/10.1007/JHEP05(2025)155} {\bibfield  {journal} {\bibinfo  {journal} {JHEP}\ }\textbf {\bibinfo {volume} {05}},\ \bibinfo {pages} {155}},\ \Eprint {https://arxiv.org/abs/2412.16149} {arXiv:2412.16149 [hep-th]} \BibitemShut {NoStop}%
\bibitem [{\citenamefont {Donnay}\ and\ \citenamefont {Herfray}(2026)}]{Donnay:2026urd}%
  \BibitemOpen
  \bibfield  {author} {\bibinfo {author} {\bibfnamefont {L.}~\bibnamefont {Donnay}}\ and\ \bibinfo {author} {\bibfnamefont {Y.}~\bibnamefont {Herfray}},\ }\bibfield  {title} {\bibinfo {title} {Infrared physics of qed and gravity from representation theory},\ }\href@noop {} {\  (\bibinfo {year} {2026})},\ \Eprint {https://arxiv.org/abs/2603.06297} {arXiv:2603.06297 [hep-th]} \BibitemShut {NoStop}%
\bibitem [{\citenamefont {Choi}\ and\ \citenamefont {Akhoury}(2018)}]{Choi:2017ylo}%
  \BibitemOpen
  \bibfield  {author} {\bibinfo {author} {\bibfnamefont {S.}~\bibnamefont {Choi}}\ and\ \bibinfo {author} {\bibfnamefont {R.}~\bibnamefont {Akhoury}},\ }\bibfield  {title} {\bibinfo {title} {{BMS Supertranslation Symmetry Implies Faddeev-Kulish Amplitudes}},\ }\href {https://doi.org/10.1007/JHEP02(2018)171} {\bibfield  {journal} {\bibinfo  {journal} {JHEP}\ }\textbf {\bibinfo {volume} {02}},\ \bibinfo {pages} {171}},\ \Eprint {https://arxiv.org/abs/1712.04551} {arXiv:1712.04551 [hep-th]} \BibitemShut {NoStop}%
\bibitem [{\citenamefont {Prabhu}\ \emph {et~al.}(2022)\citenamefont {Prabhu}, \citenamefont {Satishchandran},\ and\ \citenamefont {Wald}}]{Prabhu:2022zcr}%
  \BibitemOpen
  \bibfield  {author} {\bibinfo {author} {\bibfnamefont {K.}~\bibnamefont {Prabhu}}, \bibinfo {author} {\bibfnamefont {G.}~\bibnamefont {Satishchandran}},\ and\ \bibinfo {author} {\bibfnamefont {R.~M.}\ \bibnamefont {Wald}},\ }\bibfield  {title} {\bibinfo {title} {{Infrared finite scattering theory in quantum field theory and quantum gravity}},\ }\href {https://doi.org/10.1103/PhysRevD.106.066005} {\bibfield  {journal} {\bibinfo  {journal} {Phys. Rev. D}\ }\textbf {\bibinfo {volume} {106}},\ \bibinfo {pages} {066005} (\bibinfo {year} {2022})},\ \Eprint {https://arxiv.org/abs/2203.14334} {arXiv:2203.14334 [hep-th]} \BibitemShut {NoStop}%
\bibitem [{\citenamefont {Prabhu}\ and\ \citenamefont {Satishchandran}(2024{\natexlab{a}})}]{Prabhu:2024zwl}%
  \BibitemOpen
  \bibfield  {author} {\bibinfo {author} {\bibfnamefont {K.}~\bibnamefont {Prabhu}}\ and\ \bibinfo {author} {\bibfnamefont {G.}~\bibnamefont {Satishchandran}},\ }\bibfield  {title} {\bibinfo {title} {{Infrared finite scattering theory: scattering states and representations of the BMS group}},\ }\href {https://doi.org/10.1007/JHEP08(2024)055} {\bibfield  {journal} {\bibinfo  {journal} {JHEP}\ }\textbf {\bibinfo {volume} {08}},\ \bibinfo {pages} {055}},\ \Eprint {https://arxiv.org/abs/2402.00102} {arXiv:2402.00102 [hep-th]} \BibitemShut {NoStop}%
\bibitem [{\citenamefont {Prabhu}\ and\ \citenamefont {Satishchandran}(2024{\natexlab{b}})}]{Prabhu:2024lmg}%
  \BibitemOpen
  \bibfield  {author} {\bibinfo {author} {\bibfnamefont {K.}~\bibnamefont {Prabhu}}\ and\ \bibinfo {author} {\bibfnamefont {G.}~\bibnamefont {Satishchandran}},\ }\bibfield  {title} {\bibinfo {title} {{Infrared finite scattering theory: Amplitudes and soft theorems}},\ }\href {https://doi.org/10.1103/PhysRevD.110.085022} {\bibfield  {journal} {\bibinfo  {journal} {Phys. Rev. D}\ }\textbf {\bibinfo {volume} {110}},\ \bibinfo {pages} {085022} (\bibinfo {year} {2024}{\natexlab{b}})},\ \Eprint {https://arxiv.org/abs/2402.18637} {arXiv:2402.18637 [hep-th]} \BibitemShut {NoStop}%
\bibitem [{\citenamefont {Hannesdottir}\ and\ \citenamefont {Schwartz}(2020)}]{Hannesdottir:2019opa}%
  \BibitemOpen
  \bibfield  {author} {\bibinfo {author} {\bibfnamefont {H.}~\bibnamefont {Hannesdottir}}\ and\ \bibinfo {author} {\bibfnamefont {M.~D.}\ \bibnamefont {Schwartz}},\ }\bibfield  {title} {\bibinfo {title} {{$S$ -Matrix for massless particles}},\ }\href {https://doi.org/10.1103/PhysRevD.101.105001} {\bibfield  {journal} {\bibinfo  {journal} {Phys. Rev. D}\ }\textbf {\bibinfo {volume} {101}},\ \bibinfo {pages} {105001} (\bibinfo {year} {2020})},\ \Eprint {https://arxiv.org/abs/1911.06821} {arXiv:1911.06821 [hep-th]} \BibitemShut {NoStop}%
\bibitem [{\citenamefont {Hannesdottir}\ and\ \citenamefont {Schwartz}(2023)}]{Hannesdottir:2019umk}%
  \BibitemOpen
  \bibfield  {author} {\bibinfo {author} {\bibfnamefont {H.}~\bibnamefont {Hannesdottir}}\ and\ \bibinfo {author} {\bibfnamefont {M.~D.}\ \bibnamefont {Schwartz}},\ }\bibfield  {title} {\bibinfo {title} {{Finite $S$ matrix}},\ }\href {https://doi.org/10.1103/PhysRevD.107.L021701} {\bibfield  {journal} {\bibinfo  {journal} {Phys. Rev. D}\ }\textbf {\bibinfo {volume} {107}},\ \bibinfo {pages} {L021701} (\bibinfo {year} {2023})},\ \Eprint {https://arxiv.org/abs/1906.03271} {arXiv:1906.03271 [hep-th]} \BibitemShut {NoStop}%
\bibitem [{\citenamefont {Feal}\ \emph {et~al.}(2022)\citenamefont {Feal}, \citenamefont {Tarasov},\ and\ \citenamefont {Venugopalan}}]{Feal:2022iyn}%
  \BibitemOpen
  \bibfield  {author} {\bibinfo {author} {\bibfnamefont {X.}~\bibnamefont {Feal}}, \bibinfo {author} {\bibfnamefont {A.}~\bibnamefont {Tarasov}},\ and\ \bibinfo {author} {\bibfnamefont {R.}~\bibnamefont {Venugopalan}},\ }\bibfield  {title} {\bibinfo {title} {{QED as a many-body theory of worldlines: General formalism and infrared structure}},\ }\href {https://doi.org/10.1103/PhysRevD.106.056009} {\bibfield  {journal} {\bibinfo  {journal} {Phys. Rev. D}\ }\textbf {\bibinfo {volume} {106}},\ \bibinfo {pages} {056009} (\bibinfo {year} {2022})},\ \Eprint {https://arxiv.org/abs/2206.04188} {arXiv:2206.04188 [hep-th]} \BibitemShut {NoStop}%
\bibitem [{\citenamefont {Feal}\ \emph {et~al.}(2023)\citenamefont {Feal}, \citenamefont {Tarasov},\ and\ \citenamefont {Venugopalan}}]{Feal:2022ufw}%
  \BibitemOpen
  \bibfield  {author} {\bibinfo {author} {\bibfnamefont {X.}~\bibnamefont {Feal}}, \bibinfo {author} {\bibfnamefont {A.}~\bibnamefont {Tarasov}},\ and\ \bibinfo {author} {\bibfnamefont {R.}~\bibnamefont {Venugopalan}},\ }\bibfield  {title} {\bibinfo {title} {{QED as a many-body theory of worldlines. II. All-order S-matrix formalism}},\ }\href {https://doi.org/10.1103/PhysRevD.107.096021} {\bibfield  {journal} {\bibinfo  {journal} {Phys. Rev. D}\ }\textbf {\bibinfo {volume} {107}},\ \bibinfo {pages} {096021} (\bibinfo {year} {2023})},\ \Eprint {https://arxiv.org/abs/2211.15712} {arXiv:2211.15712 [hep-th]} \BibitemShut {NoStop}%
\bibitem [{\citenamefont {Arefeva}\ \emph {et~al.}(1974)\citenamefont {Arefeva}, \citenamefont {Faddeev},\ and\ \citenamefont {Slavnov}}]{Arefeva:1974jv}%
  \BibitemOpen
  \bibfield  {author} {\bibinfo {author} {\bibfnamefont {I.~Y.}\ \bibnamefont {Arefeva}}, \bibinfo {author} {\bibfnamefont {L.~D.}\ \bibnamefont {Faddeev}},\ and\ \bibinfo {author} {\bibfnamefont {A.~A.}\ \bibnamefont {Slavnov}},\ }\bibfield  {title} {\bibinfo {title} {{Generating Functional for the s Matrix in Gauge Theories}},\ }\href {https://doi.org/10.1007/BF01038094} {\bibfield  {journal} {\bibinfo  {journal} {Teor. Mat. Fiz.}\ }\textbf {\bibinfo {volume} {21}},\ \bibinfo {pages} {311} (\bibinfo {year} {1974})}\BibitemShut {NoStop}%
\bibitem [{\citenamefont {Faddeev}\ and\ \citenamefont {Slavnov}(1980)}]{Faddeev:1980be}%
  \BibitemOpen
  \bibfield  {author} {\bibinfo {author} {\bibfnamefont {L.~D.}\ \bibnamefont {Faddeev}}\ and\ \bibinfo {author} {\bibfnamefont {A.~A.}\ \bibnamefont {Slavnov}},\ }\href@noop {} {\emph {\bibinfo {title} {{Gauge fields. Introduction to quantum theory}}}},\ Vol.~\bibinfo {volume} {50}\ (\bibinfo {year} {1980})\BibitemShut {NoStop}%
\bibitem [{\citenamefont {Rosly}\ and\ \citenamefont {Selivanov}(1997)}]{Rosly:1996vr}%
  \BibitemOpen
  \bibfield  {author} {\bibinfo {author} {\bibfnamefont {A.~A.}\ \bibnamefont {Rosly}}\ and\ \bibinfo {author} {\bibfnamefont {K.~G.}\ \bibnamefont {Selivanov}},\ }\bibfield  {title} {\bibinfo {title} {{On amplitudes in selfdual sector of Yang-Mills theory}},\ }\href {https://doi.org/10.1016/S0370-2693(97)00268-2} {\bibfield  {journal} {\bibinfo  {journal} {Phys. Lett. B}\ }\textbf {\bibinfo {volume} {399}},\ \bibinfo {pages} {135} (\bibinfo {year} {1997})},\ \Eprint {https://arxiv.org/abs/hep-th/9611101} {arXiv:hep-th/9611101} \BibitemShut {NoStop}%
\bibitem [{\citenamefont {Adamo}\ \emph {et~al.}(2022)\citenamefont {Adamo}, \citenamefont {Cristofoli},\ and\ \citenamefont {Tourkine}}]{Adamo:2021rfq}%
  \BibitemOpen
  \bibfield  {author} {\bibinfo {author} {\bibfnamefont {T.}~\bibnamefont {Adamo}}, \bibinfo {author} {\bibfnamefont {A.}~\bibnamefont {Cristofoli}},\ and\ \bibinfo {author} {\bibfnamefont {P.}~\bibnamefont {Tourkine}},\ }\bibfield  {title} {\bibinfo {title} {{Eikonal amplitudes from curved backgrounds}},\ }\href {https://doi.org/10.21468/SciPostPhys.13.2.032} {\bibfield  {journal} {\bibinfo  {journal} {SciPost Phys.}\ }\textbf {\bibinfo {volume} {13}},\ \bibinfo {pages} {032} (\bibinfo {year} {2022})},\ \Eprint {https://arxiv.org/abs/2112.09113} {arXiv:2112.09113 [hep-th]} \BibitemShut {NoStop}%
\bibitem [{\citenamefont {Lipinski~Jusinskas}(2026)}]{LipinskiJusinskas:2026ctz}%
  \BibitemOpen
  \bibfield  {author} {\bibinfo {author} {\bibfnamefont {R.}~\bibnamefont {Lipinski~Jusinskas}},\ }\bibfield  {title} {\bibinfo {title} {{Perturbiner methods in scattering amplitude}},\ }\href@noop {} {\  (\bibinfo {year} {2026})},\ \Eprint {https://arxiv.org/abs/2607.06661} {arXiv:2607.06661 [hep-th]} \BibitemShut {NoStop}%
\bibitem [{\citenamefont {Jain}\ \emph {et~al.}(2026)\citenamefont {Jain}, \citenamefont {Kundu}, \citenamefont {Minwalla}, \citenamefont {Parrikar}, \citenamefont {Prabhu},\ and\ \citenamefont {Shrivastava}}]{Jain:2023fxc}%
  \BibitemOpen
  \bibfield  {author} {\bibinfo {author} {\bibfnamefont {D.}~\bibnamefont {Jain}}, \bibinfo {author} {\bibfnamefont {S.}~\bibnamefont {Kundu}}, \bibinfo {author} {\bibfnamefont {S.}~\bibnamefont {Minwalla}}, \bibinfo {author} {\bibfnamefont {O.}~\bibnamefont {Parrikar}}, \bibinfo {author} {\bibfnamefont {S.~G.}\ \bibnamefont {Prabhu}},\ and\ \bibinfo {author} {\bibfnamefont {P.}~\bibnamefont {Shrivastava}},\ }\bibfield  {title} {\bibinfo {title} {{The S-matrix and boundary correlators in flat space}},\ }\href {https://doi.org/10.1007/JHEP02(2026)151} {\bibfield  {journal} {\bibinfo  {journal} {JHEP}\ }\textbf {\bibinfo {volume} {02}},\ \bibinfo {pages} {151}},\ \Eprint {https://arxiv.org/abs/2311.03443} {arXiv:2311.03443 [hep-th]} \BibitemShut {NoStop}%
\bibitem [{\citenamefont {Kim}\ \emph {et~al.}(2023)\citenamefont {Kim}, \citenamefont {Kraus}, \citenamefont {Monten},\ and\ \citenamefont {Myers}}]{Kim:2023qbl}%
  \BibitemOpen
  \bibfield  {author} {\bibinfo {author} {\bibfnamefont {S.}~\bibnamefont {Kim}}, \bibinfo {author} {\bibfnamefont {P.}~\bibnamefont {Kraus}}, \bibinfo {author} {\bibfnamefont {R.}~\bibnamefont {Monten}},\ and\ \bibinfo {author} {\bibfnamefont {R.~M.}\ \bibnamefont {Myers}},\ }\bibfield  {title} {\bibinfo {title} {{S-matrix path integral approach to symmetries and soft theorems}},\ }\href {https://doi.org/10.1007/JHEP10(2023)036} {\bibfield  {journal} {\bibinfo  {journal} {JHEP}\ }\textbf {\bibinfo {volume} {10}},\ \bibinfo {pages} {036}},\ \Eprint {https://arxiv.org/abs/2307.12368} {arXiv:2307.12368 [hep-th]} \BibitemShut {NoStop}%
\bibitem [{\citenamefont {Kraus}\ and\ \citenamefont {Myers}(2025{\natexlab{a}})}]{Kraus:2024gso}%
  \BibitemOpen
  \bibfield  {author} {\bibinfo {author} {\bibfnamefont {P.}~\bibnamefont {Kraus}}\ and\ \bibinfo {author} {\bibfnamefont {R.~M.}\ \bibnamefont {Myers}},\ }\bibfield  {title} {\bibinfo {title} {{Carrollian partition functions and the flat limit of AdS}},\ }\href {https://doi.org/10.1007/JHEP01(2025)183} {\bibfield  {journal} {\bibinfo  {journal} {JHEP}\ }\textbf {\bibinfo {volume} {01}},\ \bibinfo {pages} {183}},\ \Eprint {https://arxiv.org/abs/2407.13668} {arXiv:2407.13668 [hep-th]} \BibitemShut {NoStop}%
\bibitem [{\citenamefont {Kraus}\ and\ \citenamefont {Myers}(2025{\natexlab{b}})}]{Kraus:2025wgi}%
  \BibitemOpen
  \bibfield  {author} {\bibinfo {author} {\bibfnamefont {P.}~\bibnamefont {Kraus}}\ and\ \bibinfo {author} {\bibfnamefont {R.~M.}\ \bibnamefont {Myers}},\ }\bibfield  {title} {\bibinfo {title} {{Carrollian partition function for bulk Yang-Mills theory}},\ }\href {https://doi.org/10.1007/JHEP08(2025)180} {\bibfield  {journal} {\bibinfo  {journal} {JHEP}\ }\textbf {\bibinfo {volume} {08}},\ \bibinfo {pages} {180}},\ \Eprint {https://arxiv.org/abs/2503.00916} {arXiv:2503.00916 [hep-th]} \BibitemShut {NoStop}%
\bibitem [{\citenamefont {Isen}\ \emph {et~al.}(2026)\citenamefont {Isen}, \citenamefont {Kraus}, \citenamefont {Monten},\ and\ \citenamefont {Myers}}]{Isen:2026xoc}%
  \BibitemOpen
  \bibfield  {author} {\bibinfo {author} {\bibfnamefont {J.}~\bibnamefont {Isen}}, \bibinfo {author} {\bibfnamefont {P.}~\bibnamefont {Kraus}}, \bibinfo {author} {\bibfnamefont {R.}~\bibnamefont {Monten}},\ and\ \bibinfo {author} {\bibfnamefont {R.~M.}\ \bibnamefont {Myers}},\ }\bibfield  {title} {\bibinfo {title} {{The gravitational S-matrix from the path integral: asymptotic symmetries and soft theorems}},\ }\href@noop {} {\  (\bibinfo {year} {2026})},\ \Eprint {https://arxiv.org/abs/2603.17045} {arXiv:2603.17045 [hep-th]} \BibitemShut {NoStop}%
\bibitem [{\citenamefont {Ammon}\ \emph {et~al.}(2026)\citenamefont {Ammon}, \citenamefont {Capone},\ and\ \citenamefont {Sieling}}]{Ammon:2025jmy}%
  \BibitemOpen
  \bibfield  {author} {\bibinfo {author} {\bibfnamefont {M.}~\bibnamefont {Ammon}}, \bibinfo {author} {\bibfnamefont {F.}~\bibnamefont {Capone}},\ and\ \bibinfo {author} {\bibfnamefont {C.}~\bibnamefont {Sieling}},\ }\bibfield  {title} {\bibinfo {title} {{Flat holography {\&} holographic renormalization: scalar field}},\ }\href {https://doi.org/10.1007/JHEP07(2026)124} {\bibfield  {journal} {\bibinfo  {journal} {JHEP}\ }\textbf {\bibinfo {volume} {07}},\ \bibinfo {pages} {124}},\ \Eprint {https://arxiv.org/abs/2512.14818} {arXiv:2512.14818 [hep-th]} \BibitemShut {NoStop}%
\bibitem [{\citenamefont {Hirai}\ and\ \citenamefont {Sugishita}(2023)}]{Hirai:2022yqw}%
  \BibitemOpen
  \bibfield  {author} {\bibinfo {author} {\bibfnamefont {H.}~\bibnamefont {Hirai}}\ and\ \bibinfo {author} {\bibfnamefont {S.}~\bibnamefont {Sugishita}},\ }\bibfield  {title} {\bibinfo {title} {{Dress code for infrared safe scattering in QED}},\ }\href {https://doi.org/10.1093/ptep/ptad057} {\bibfield  {journal} {\bibinfo  {journal} {PTEP}\ }\textbf {\bibinfo {volume} {2023}},\ \bibinfo {pages} {053B04} (\bibinfo {year} {2023})},\ \Eprint {https://arxiv.org/abs/2209.00608} {arXiv:2209.00608 [hep-th]} \BibitemShut {NoStop}%
\bibitem [{Note2()}]{Note2}%
  \BibitemOpen
  \bibinfo {note} {Note that the literature, including the original computation in \cite {Kulish:1970ut}, uses mostly $:\protect \hat {\rho }(\protect \mathbf p) \protect \hat {\rho }(\protect \mathbf q):$ instead, the difference being self-energy terms. This means we assume here instead that UV divergences in the theory have been renormalized, but without getting rid of the IR divergences of the self-energy diagrams. Note, however, that the formalism works either way.}\BibitemShut {Stop}%
\bibitem [{\citenamefont {Chung}(1965)}]{Chung:1965zza}%
  \BibitemOpen
  \bibfield  {author} {\bibinfo {author} {\bibfnamefont {V.}~\bibnamefont {Chung}},\ }\bibfield  {title} {\bibinfo {title} {{Infrared Divergence in Quantum Electrodynamics}},\ }\href {https://doi.org/10.1103/PhysRev.140.B1110} {\bibfield  {journal} {\bibinfo  {journal} {Phys. Rev.}\ }\textbf {\bibinfo {volume} {140}},\ \bibinfo {pages} {B1110} (\bibinfo {year} {1965})}\BibitemShut {NoStop}%
\bibitem [{Note3()}]{Note3}%
  \BibitemOpen
  \bibinfo {note} {When polarization or spin indices are omitted, the derivatives or generators of the Heisenberg algebra correspond to either the photon or the fermion algebra.}\BibitemShut {Stop}%
\bibitem [{Note4()}]{Note4}%
  \BibitemOpen
  \bibinfo {note} {Note that these differential displacement operators depend on $\protect \bar \alpha _\mu $; calling this also a displacement operator is subject to notation.}\BibitemShut {Stop}%
\bibitem [{\citenamefont {Weinberg}(1965)}]{Weinberg:1965nx}%
  \BibitemOpen
  \bibfield  {author} {\bibinfo {author} {\bibfnamefont {S.}~\bibnamefont {Weinberg}},\ }\bibfield  {title} {\bibinfo {title} {{Infrared photons and gravitons}},\ }\href {https://doi.org/10.1103/PhysRev.140.B516} {\bibfield  {journal} {\bibinfo  {journal} {Phys. Rev.}\ }\textbf {\bibinfo {volume} {140}},\ \bibinfo {pages} {B516} (\bibinfo {year} {1965})}\BibitemShut {NoStop}%
\bibitem [{Note5()}]{Note5}%
  \BibitemOpen
  \bibinfo {note} {In the language of non-cyclic geometric phases, the Coulomb phase is therefore a dynamical phase \cite {PhysRevA.52.2576}. For the normalized photon coherent states one may write it in terms of the Berry connection one-form associated with the cloud photon coherent state space. For example in the quantum-mechanical case, $A = i{}_N \langle z|d|z \rangle _N$, where $\mathinner {|{z}\rangle }_N$ is the normalised coherent state in \protect \eqref {eq:defcoherent} and $d$ the exterior derivative. With our sign convention, $\Phi = -\DOTSI \intop \ilimits@ A$.}\BibitemShut {Stop}%
\bibitem [{Note6()}]{Note6}%
  \BibitemOpen
  \bibinfo {note} {Note that this differs from the non-relativistic case, where only the Coulomb phase contributes.}\BibitemShut {Stop}%
\bibitem [{\citenamefont {Lippstreu}(2025)}]{Lippstreu:2025jit}%
  \BibitemOpen
  \bibfield  {author} {\bibinfo {author} {\bibfnamefont {L.}~\bibnamefont {Lippstreu}},\ }\href@noop {} {\bibinfo {title} {{Analytic Properties of Infrared-Finite Amplitudes in Theories with Long-Range Forces}}} (\bibinfo {year} {2025}),\ \Eprint {https://arxiv.org/abs/2505.04702} {arXiv:2505.04702 [hep-th]} \BibitemShut {NoStop}%
\bibitem [{Note7()}]{Note7}%
  \BibitemOpen
  \bibinfo {note} {Furthermore, in gravity, the Coulomb phase is related to classical observables such as the Shapiro time delay.}\BibitemShut {Stop}%
\bibitem [{Note8()}]{Note8}%
  \BibitemOpen
  \bibinfo {note} {The factors we discarded in section \protect \eqref {sec:pathint} to stay at tree level are diagrammatically represented by loops with integration over soft momenta. The normalizations $\protect \mathcal {N}_{\protect \text {ii}}$, $\protect \mathcal {N}_{\protect \text {oo}}$ and $\protect \mathcal {N}_{\protect \text {oi}}$ contribute photon lines that connect different in- and out-going clouds. The Coulomb-phase is usually denoted as a line between two Fermion legs and the shift corresponds to the remaining possible processes, a soft photon line connecting a cloud with the bulk QED graph.}\BibitemShut {Stop}%
\bibitem [{\citenamefont {Choi}\ and\ \citenamefont {Mitra}(2026{\natexlab{a}})}]{Choi:2026swo}%
  \BibitemOpen
  \bibfield  {author} {\bibinfo {author} {\bibfnamefont {S.}~\bibnamefont {Choi}}\ and\ \bibinfo {author} {\bibfnamefont {P.}~\bibnamefont {Mitra}},\ }\bibfield  {title} {\bibinfo {title} {{Dressed Fock Spaces in Gauge Theory and Gravity}},\ }\href@noop {} {\  (\bibinfo {year} {2026}{\natexlab{a}})},\ \Eprint {https://arxiv.org/abs/2606.15988} {arXiv:2606.15988 [hep-th]} \BibitemShut {NoStop}%
\bibitem [{\citenamefont {Choi}\ and\ \citenamefont {Akhoury}(2019)}]{Choi:2019rlz}%
  \BibitemOpen
  \bibfield  {author} {\bibinfo {author} {\bibfnamefont {S.}~\bibnamefont {Choi}}\ and\ \bibinfo {author} {\bibfnamefont {R.}~\bibnamefont {Akhoury}},\ }\bibfield  {title} {\bibinfo {title} {{Subleading soft dressings of asymptotic states in QED and perturbative quantum gravity}},\ }\href {https://doi.org/10.1007/JHEP09(2019)031} {\bibfield  {journal} {\bibinfo  {journal} {JHEP}\ }\textbf {\bibinfo {volume} {09}},\ \bibinfo {pages} {031}},\ \Eprint {https://arxiv.org/abs/1907.05438} {arXiv:1907.05438 [hep-th]} \BibitemShut {NoStop}%
\bibitem [{\citenamefont {Christodoulou}\ and\ \citenamefont {Toumbas}(2026)}]{Christodoulou:2026gvt}%
  \BibitemOpen
  \bibfield  {author} {\bibinfo {author} {\bibfnamefont {S.}~\bibnamefont {Christodoulou}}\ and\ \bibinfo {author} {\bibfnamefont {N.}~\bibnamefont {Toumbas}},\ }\bibfield  {title} {\bibinfo {title} {{Subleading soft dressings for QED scattering states}},\ }\href@noop {} {\  (\bibinfo {year} {2026})},\ \Eprint {https://arxiv.org/abs/2603.18587} {arXiv:2603.18587 [hep-th]} \BibitemShut {NoStop}%
\bibitem [{\citenamefont {Choi}\ and\ \citenamefont {Mitra}(2026{\natexlab{b}})}]{Choi:2026dhz}%
  \BibitemOpen
  \bibfield  {author} {\bibinfo {author} {\bibfnamefont {S.}~\bibnamefont {Choi}}\ and\ \bibinfo {author} {\bibfnamefont {P.}~\bibnamefont {Mitra}},\ }\bibfield  {title} {\bibinfo {title} {{A Holographic Model for Soft Photons and Gravitons in Four Dimensions}},\ }\href@noop {} {\  (\bibinfo {year} {2026}{\natexlab{b}})},\ \Eprint {https://arxiv.org/abs/2603.14499} {arXiv:2603.14499 [hep-th]} \BibitemShut {NoStop}%
\bibitem [{\citenamefont {Kapec}(2024)}]{Kapec:2022hih}%
  \BibitemOpen
  \bibfield  {author} {\bibinfo {author} {\bibfnamefont {D.}~\bibnamefont {Kapec}},\ }\bibfield  {title} {\bibinfo {title} {{Soft particles and infinite-dimensional geometry}},\ }\href {https://doi.org/10.1088/1361-6382/ad0514} {\bibfield  {journal} {\bibinfo  {journal} {Class. Quant. Grav.}\ }\textbf {\bibinfo {volume} {41}},\ \bibinfo {pages} {015001} (\bibinfo {year} {2024})},\ \Eprint {https://arxiv.org/abs/2210.00606} {arXiv:2210.00606 [hep-th]} \BibitemShut {NoStop}%
\bibitem [{\citenamefont {Jevicki}\ and\ \citenamefont {Lee}(1988)}]{Jevicki:1987ax}%
  \BibitemOpen
  \bibfield  {author} {\bibinfo {author} {\bibfnamefont {A.}~\bibnamefont {Jevicki}}\ and\ \bibinfo {author} {\bibfnamefont {C.-k.}\ \bibnamefont {Lee}},\ }\bibfield  {title} {\bibinfo {title} {{The S Matrix Generating Functional and Effective Action}},\ }\href {https://doi.org/10.1103/PhysRevD.37.1485} {\bibfield  {journal} {\bibinfo  {journal} {Phys. Rev. D}\ }\textbf {\bibinfo {volume} {37}},\ \bibinfo {pages} {1485} (\bibinfo {year} {1988})}\BibitemShut {NoStop}%
\bibitem [{\citenamefont {Pati}(1995)}]{PhysRevA.52.2576}%
  \BibitemOpen
  \bibfield  {author} {\bibinfo {author} {\bibfnamefont {A.~K.}\ \bibnamefont {Pati}},\ }\bibfield  {title} {\bibinfo {title} {Geometric aspects of noncyclic quantum evolutions},\ }\href {https://doi.org/10.1103/PhysRevA.52.2576} {\bibfield  {journal} {\bibinfo  {journal} {Phys. Rev. A}\ }\textbf {\bibinfo {volume} {52}},\ \bibinfo {pages} {2576} (\bibinfo {year} {1995})}\BibitemShut {NoStop}%
\end{thebibliography}%
